\documentclass[aps,prl,twocolumn,longbibliography,superscriptaddress,notitlepage,floatfix]{revtex4-2}
\renewcommand{\figurename}{Fig.}
\usepackage{amsmath,amsfonts,amssymb,color,epsfig,graphics,graphicx,latexsym,theorem,url,multirow,diagbox}
\usepackage[colorlinks,citecolor=blue,linkcolor=blue,urlcolor=blue]{hyperref}
\usepackage[utf8]{inputenc}
\usepackage{booktabs}
\usepackage[T1]{fontenc}
\usepackage{courier}
\usepackage{listings}
\usepackage{latexsym}
\usepackage{balance}
\usepackage{dcolumn}
\usepackage{comment}
\usepackage{times} 
\usepackage{ulem}
\usepackage{xcolor}
\usepackage{color}
\usepackage{pifont} 
\usepackage{newfloat}
\definecolor{darkgreenRGB}{RGB}{0, 180, 0}

\usepackage{soul}
\soulregister\cite7 
\soulregister\citep7 
\soulregister\citet7 
\soulregister\ref7 
\soulregister\pageref7 
\soulregister\item7
\soulregister\eqref7

\usepackage{hyperref}

\usepackage{braket}
\usepackage{bm}

\begin{document}

\title{Experimental realization of the bucket-brigade quantum random access memory}

\affiliation{School of Physics, ZJU-Hangzhou Global Scientific and Technological Innovation Center, and Zhejiang Key Laboratory of Micro-nano Quantum Chips and Quantum Control, Zhejiang University, Hangzhou, China\\
$^{2}$ College of Computer Science, and ZJU-Ningbo Global Innovation Center, Zhejiang University, Hangzhou, China \\
$^{3}$ China Mobile (Suzhou) Software Technology Co., Ltd., Suzhou 215163, China
}

\author{Fanhao Shen$^{1}$}\thanks{These authors contributed equally to this work.}
\author{Yujie Ji$^{1}$}\thanks{These authors contributed equally to this work.}
\author{Debin Xiang$^{2}$}\thanks{These authors contributed equally to this work.}
\author{Yanzhe Wang$^{1}$}
\author{Ke Wang$^{1}$}
\author{Chuanyu Zhang$^{1}$}
\author{Aosai Zhang$^{1}$}
\author{Yiren Zou$^{1}$}
\author{Yu Gao$^{1}$}
\author{Zhengyi Cui$^{1}$}
\author{Gongyu Liu$^{1}$}
\author{Jianan Yang$^{1}$}
\author{Yihang Han$^{1}$}
\author{Jinfeng Deng$^{1}$}
\author{Anbang Wang$^{3}$}
\author{Zhihong Zhang$^{3}$}
\author{Hekang Li$^{1}$}
\author{Qiujiang Guo$^{1}$}
\author{Pengfei Zhang$^{1}$}
\author{Chao Song$^{1}$}\email{chaosong@zju.edu.cn}
\author{Liqiang Lu$^{2}$}\email{ liqianglu@zju.edu.cn}
\author{Zhen Wang$^{1}$}\email{2010wangzhen@zju.edu.cn}
\author{Jianwei Yin$^{2}$}\email{zjuyjw@cs.zju.edu.cn}

\begin{abstract}
Quantum random access memory (QRAM) enables efficient classical data access for quantum computers -- a prerequisite for many quantum algorithms to achieve quantum speedup.
Despite various proposals, the experimental realization of QRAM remains largely unexplored.
Here, we experimentally investigate the circuit-based bucket-brigade QRAM with a superconducting quantum processor.
To facilitate the experimental implementation,
we introduce a hardware-efficient gate decomposition scheme for quantum routers, which effectively reduces the depth of the QRAM circuit by more than $30\%$ compared to the conventional controlled-SWAP-based implementation.
We further propose an error mitigation method to boost the QRAM query fidelity.
With these techniques, we are able to experimentally implement the QRAM architectures with two and three layers, achieving query fidelities up to 0.800 $\pm$ 0.026 and 0.604$\pm$0.005, respectively.
Additionally, we study the error propagation mechanism and the scalability of our QRAM implementation, providing experimental evidence for the noise resilience nature of the bucket-brigade QRAM architecture.
Our results highlight the potential of superconducting quantum processors for realizing a scalable QRAM architecture.

\end{abstract}

\maketitle

\vspace{.6cm}
\noindent
The quantum query of classical data is a fundamental requirement for numerous quantum algorithms such as quantum machine learning~\cite{gupta2001quantum,rebentrost2014quantum,PhysRevA.99.012326,10130283,Liu2024,rspa}, quantum simulation~\cite{Berry_2019,PhysRevX.8.041015,Childs2022quantumsimulationof,PhysRevA.110.032439,Feniou2024}, Grover's algorithm~\cite{Grover}, and a host of other areas~\cite{PhysRevLett.100.230502,PhysRevLett.109.050505,Lloyd2016,PhysRevLett.103.150502}.
In practice, this querying process can often dominate the computational overhead, significantly impacting the potential quantum advantage~\cite{aaronson2015read}.
Quantum random access memory (QRAM)~\cite{giovannetti2008quantum,giovannetti2008architectures,di2020fault,hann2021resilience,paler2020parallelizing,xu2023systems,hann2019hardware,chen2023efficient,wang2024fundamental,arunachalam2015robustness,Bugalho2023,Yongshan2025,PRXQuantum.2.030319} 
offers a promising solution to this dilemma, which allows quantum processors to access and retrieve data in superposition via quantum addresses. 
Considering a memory of size $N$ where each classical data value $x_{i}$ is stored at address $i \in \{0,1,\cdots,N-1\}$, the QRAM performs the following unitary transformation:
\begin{equation} \label{equ:qram}
    \sum_{i=0}^{N-1} \alpha_{i} \ket{i}_\text{A} \ket{0}_\text{D} \overset{\text{QRAM}}{\xrightarrow{\hspace{1cm}}}
    \sum_{i=0}^{N-1} \alpha_{i} \ket{i}_\text{A} \ket{x_{i}}_\text{D},
\end{equation}
where $\alpha_i$ is the amplitude of each address in the superposition, and $\ket{\cdot}_\text{A}(\ket{\cdot}_\text{D})$ is the address (data) qubit register storing the input (output). 
Notably, the above operation can be realized with a time complexity of $\mathcal{O}(log N)$, which is optimal for implementing any oracle-based quantum algorithm~\cite{arunachalam2015robustness}.

Among various QRAM architecture designs, the bucket-brigade QRAM architecture~\cite{giovannetti2008quantum} turns out to be a promising approach due to its logarithmic infidelity scaling with the memory size~\cite{hann2021resilience}.
The architecture consists of a binary tree of quantum routers, each capable of coherently directing the input signal to superpositions of output paths based on quantum addresses~\cite{wang2024quantumrandomaccessmemory}.
Such quantum routers essentially involve four-body interactions, making their physical implementation challenging.
Despite various theoretical proposals~\cite{hann2019hardware, PRXQuantum.5.020312,PRXQuantum.2.030319}, experimental progress was limited until recent breakthroughs demonstrating deterministic quantum routers on superconducting platforms~\cite{miao2025implementation,zhang2025quantumrouters}.
Nevertheless, a scalable implementation of the full bucket-brigade QRAM architecture, along with a comprehensive understanding of its error propagation under realistic noise conditions, remains largely unexplored.

\begin{figure*}[t]
\centering
\includegraphics[width=0.85\textwidth]{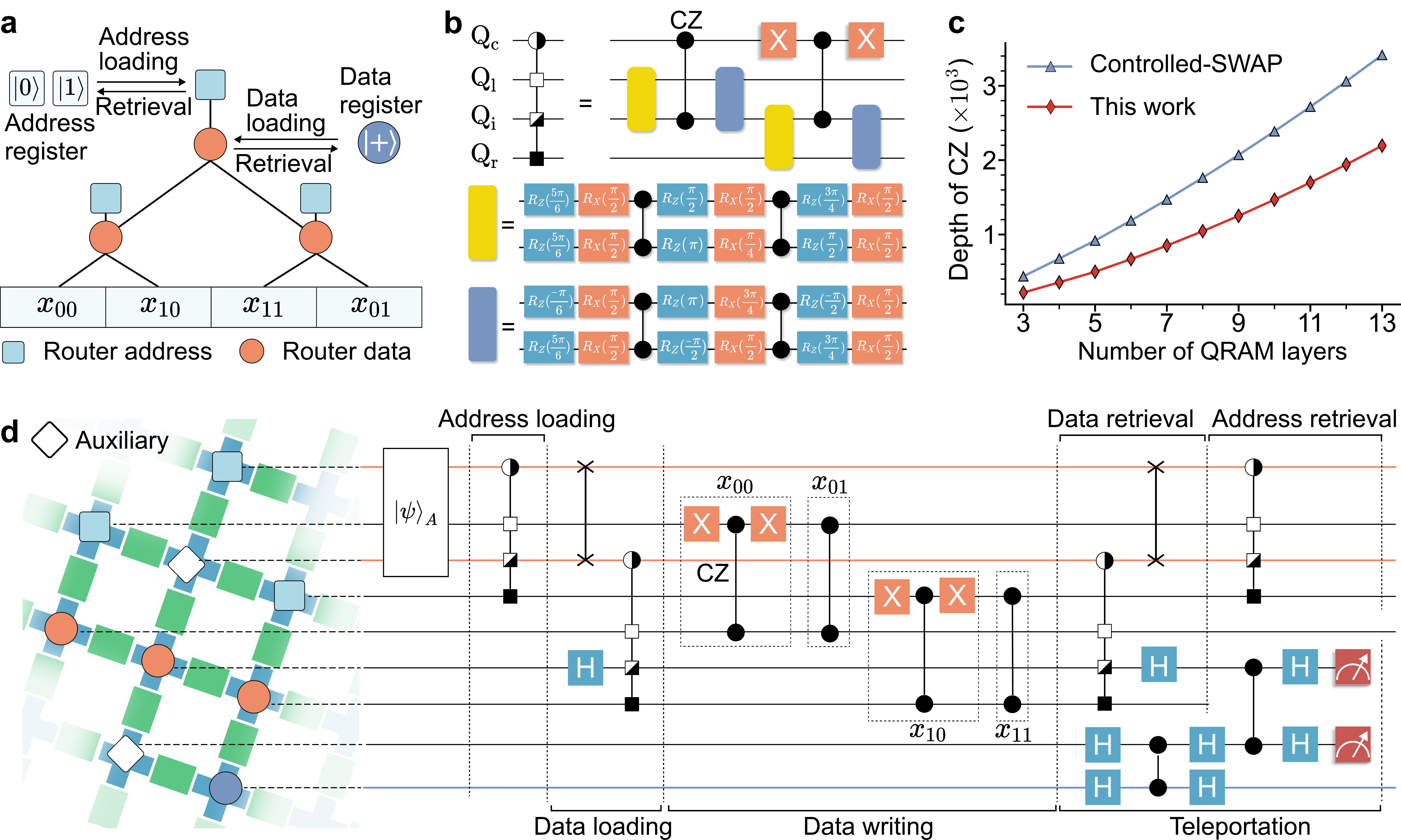}
\caption{
\textbf{Experimental implementation of QRAM with a superconducting quantum processor.} \textbf{a}, Schematic of the two-layer bucket-brigade QRAM, which can query four classical data bits, denoted as $x_{00}$, $x_{01}$, $x_{10}$, and $x_{11}$, according to the query address.
The qubits in the QRAM routers can be divided into two groups, one for storing the address information, and the other for loading the data bus.
The data bus encode the data bit in the $\ket{\pm}$ basis to ensure that the data writing operation acts trivially on all qubits that do not contain the data bus -- specifically, those remaining in the $\ket{0}$ state.
\textbf{b}, The experimental circuit for implementing the quantum routing operation when the incident qubit ($\text{Q}_\text{i}$) is connected to the control ($\text{Q}_\text{c}$) and two output ($\text{Q}_\text{l}$ and $\text{Q}_\text{r}$) qubits simultaneously.
\textbf{c}, Comparison of CZ gate depths in the QRAM circuit across varying numbers of layers, evaluated for the conventional controlled-SWAP-based approach and our proposed hardware-efficient decomposition scheme, respectively.
Here we consider the implementation based on the H-tree recursive mapping and quantum teleportation techniques. 
\textbf{d}, Schematic of the device layout and the corresponding quantum circuit of the two-layer bucket-brigade QRAM. 
For simplicity, we reuse the router address qubit in the first layer and one the auxiliary qubits as the address registers before and after the QRAM operation.
After initializing the address registers in $\ket{\psi}_\text{A}$, the circuit sequentially conducts standard QRAM operations including address loading, data loading, data writing, data retrieval, and address retrieval.
The gates in the dashed boxes during data writing are applied conditional on the state of the classical data bits $x_{00}$, $x_{01}$, $x_{10}$, and $x_{11}$, respectively.
}
\label{fig1}
\end{figure*}

Here, we report the experimental realization and investigation of a scalable bucket-brigade QRAM architecture on a high-performance programmable superconducting quantum processor, where qubits are arranged in a two-dimensional (2D) square grid.
By exploring the inherent degree of freedom in QRAM, we propose a hardware-efficient gate decomposition scheme of the quantum routing operation, which reduces the circuit depth by more than $30\%$ compared to the traditional routing circuit based on controlled-SWAP (CSWAP) gate~\cite{hann2019hardware}.
Besides, our implementation integrates quantum teleportation~\cite{Hu2023}, a key technique to maintain the logarithmic time complexity of QRAM during its mapping to the 2D grid of our device.
With these techniques, we present an end-to-end demonstration of the bucket-brigade QRAM architecture consisting of a full cycle of address loading, data loading, data writing, data retrieval, and address retrieval.
We begin by analyzing a two-layer QRAM (\autoref{fig1}a), where we comprehensively investigate its performance with different quantum addresses and classical data configurations, observing the correlation between the query fidelity and the number of address components.
By utilizing router qubits as error indicators, we develop an efficient error mitigation technique to enhance the query fidelity.
We then move on to a three-layer QRAM, where we are able to experimentally validate the localized nature of error propagation in QRAM through observing its performance with varying errors on different branches.
Furthermore, by capturing the entanglement structure among the memory components, we experimentally identify the origin of the noise resilience of the bucket-brigade QRAM architecture. 
We finish by providing the scalability analysis of our QRAM implementation based on gate error models.

\vspace{.6cm}
\noindent\textbf{\large{}Experimental implementation}\\
\noindent
The digital implementation of quantum routers, which involves decomposing each router into native-gate quantum circuits that respect device connectivity constraints, is highly resource-intensive.
Conventionally, a quantum router can be constructed using two CSWAP gates~\cite{hann2021resilience}, with each CSWAP requiring 8 or 10 CZ gates depending on the connectivity between the control qubit and target qubits~\cite{mq2024shallow}.
During QRAM operations, routing is confined to specific subspaces, exposing additional circuit optimization opportunities (see Supplementary Information for details).
By identifying a family of functionally equivalent unitaries for the routing operation, we select the most efficient variant when decomposed into the native gate set of arbitrary single-qubit gates and CZ gates between nearest-neighboring qubits.
Our optimized implementation, shown in \autoref{fig1}b, demonstrates a particularly efficient case where the incident qubit ($\text{Q}_\text{i}$) connects directly to the control ($\text{Q}_\text{c}$) and two output ($\text{Q}_\text{l}$ and $\text{Q}_\text{r}$) qubits, requiring only 10 CZ gates -- a $37.5\%$ reduction compared to the CSWAP-based approach.
For other connectivity configurations, corresponding circuits are provided in the Supplementary Information.
Overall, our implementation achieves more than a $30\%$ reduction in CZ gate depth, as illustrated in \autoref{fig1}c.

\begin{figure*}[t]
\centering
\includegraphics[width=0.85\textwidth]{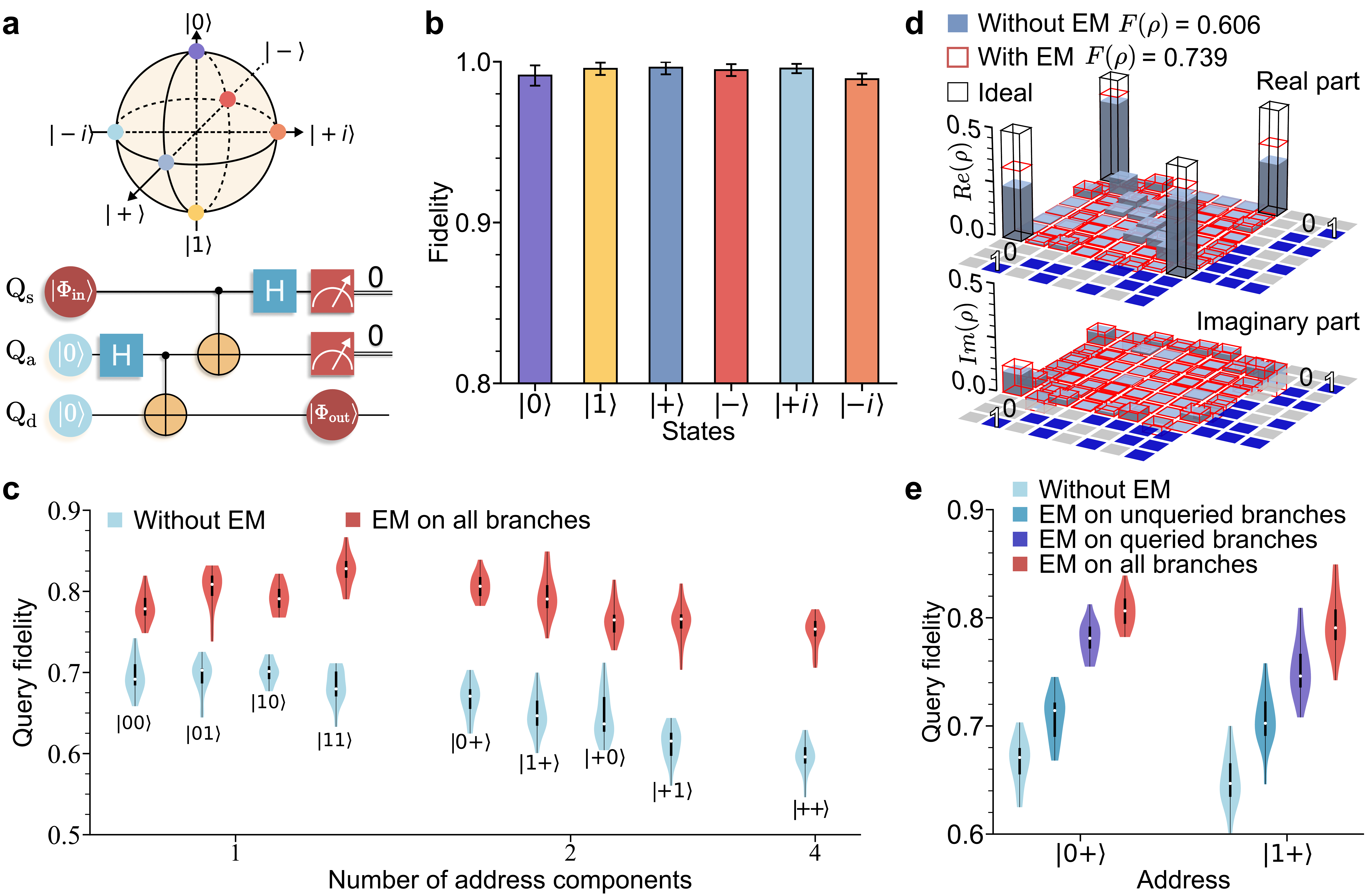}
\vspace{-.4cm}
\caption{
\textbf{Experimental quantum teleportation and QRAM performances.} \textbf{a}, The six quantum states used in the teleportation experiment and the quantum circuit of teleportation. The state in the source qubit ($\text{Q}_\text{s}$) is transferred to the data qubit ($\text{Q}_\text{d}$) through the ancillary qubit ($\text{Q}_\text{a}$). \textbf{b}, The teleportation fidelities of the six quantum states, calculated as $F \equiv \left( {\rm Tr} \sqrt{\rho_{\text{out}}^{1/2} \sigma_{\text{in}} \rho_{\text{out}}^{1/2}} \right)^{2}$, where $\rho_{\text{out}}$ is the experimentally obtained output state and $\sigma_{\text{in}}$ denotes the ideal input state. Error bars represent standard deviation of five repeated runs. \textbf{c}, Query fidelities of the two-layer bucket brigade QRAM for query addresses consisting of different numbers of components, with and without error mitigation (EM) techniques applied. For each query address, we display the distribution of query fidelities across all classical data configurations. \textbf{d}, Density matrices of GHZ states prepared using a two-layer bucket-brigade QRAM, with and without error mitigation techniques applied. \textbf{e}, Query fidelity distribution for different error mitigation strategies under query addresses $\ket{0+}$ and $\ket{1+}$ .
}
\label{fig2}
\end{figure*}
Since the bucket-brigade QRAM involves a binary tree topology, the next challenge is to map it to our square grid topology. 
Traditional methods like H-tree recursive mapping strategy~\cite{xu2023systems} can lead to exponential growth of the query latency with increasing numbers of QRAM layer, destroying the logarithmic time complexity of QRAM. 
The overhead stems from the exponentially increased number of ancillary qubits between adjacent layers of quantum routers.
To overcome this challenge, it is necessary to use quantum teleportation, which can establish inter-layer router connections with a constant circuit depth, thus preserving the logarithmic time complexity as evidenced in \autoref{fig1}c.

The complete quantum circuit for a two-layer bucket-brigade QRAM is depicted in \autoref{fig1}d, which comprises five operational phases including address loading, data loading, data writing, data retrieval, and address retrieval. 
After compilation, the circuit contains 41 layers of CZ gates interleaved with single-qubit gates.
In our experiment, we realize median fidelities of about $99.96\%$ and $99.7\%$ for the parallel single- and two-qubit gates, which are characterized through simultaneous cross-entropy benchmarking~\cite{Arute2019,Boixo2018} (see Supplementary Information for more details on the device performance).
We also profile the performance of the quantum teleportation operation used in our experiment (see Methods).
Specifically, we prepare six distinct input states on the Bloch sphere as shown in \autoref{fig2}a, and measure the fidelities of the corresponding output states after quantum teleportation with quantum state tomography.
\autoref{fig2}b provides the results, which yield an average fidelity of $\overline{F} = 0.994\pm0.005$, demonstrating the robustness of data transfer in our QRAM implementation.

\vspace{.6cm}
\noindent\textbf{\large{}QRAM performance}\\
\noindent
With the experimental bucket-brigade QRAM architecture established, we first investigate its performance under different configurations of the query address and classical data.
We quantify the performance using the query fidelity metric, defined as $F=\bra{\psi_{\text{ideal}}}\rho_{\text{address,data}}\ket{\psi_{\text{ideal}}}$, where $\rho_{\text{address,data}}$ is the density matrix of the data and address registers after QRAM query, and $\ket{\psi_{\text{ideal}}}$ is the corresponding ideal state. 
For a two-layer QRAM, there are four classical data bits, each of which can be accessed by one of the query addresses $\{\ket{00}, \ket{01}, \ket{10} ,\ket{11}\}$ through the corresponding QRAM path.
The classical data can also be queried in parallel by preparing the address qubits in arbitrary quantum superposition of the four base components.
In our experiment, we measure the query fidelities of the QRAM over all $16$ different classical data configurations for different query addresses, with the results summarized in \autoref{fig2}c.
By grouping the query addresses according to the number of base components involved, we observe a clear degradation of the query fidelity with an increasing number of components. 
Specifically, the query fidelity degrades from an average value of $0.694\pm0.020$ for single-component addresses, to $0.644\pm0.031$ and $0.595\pm0.019$ for two- and four-component addresses.

\begin{figure*}[t]
\centering
\includegraphics[width=0.63\textwidth]{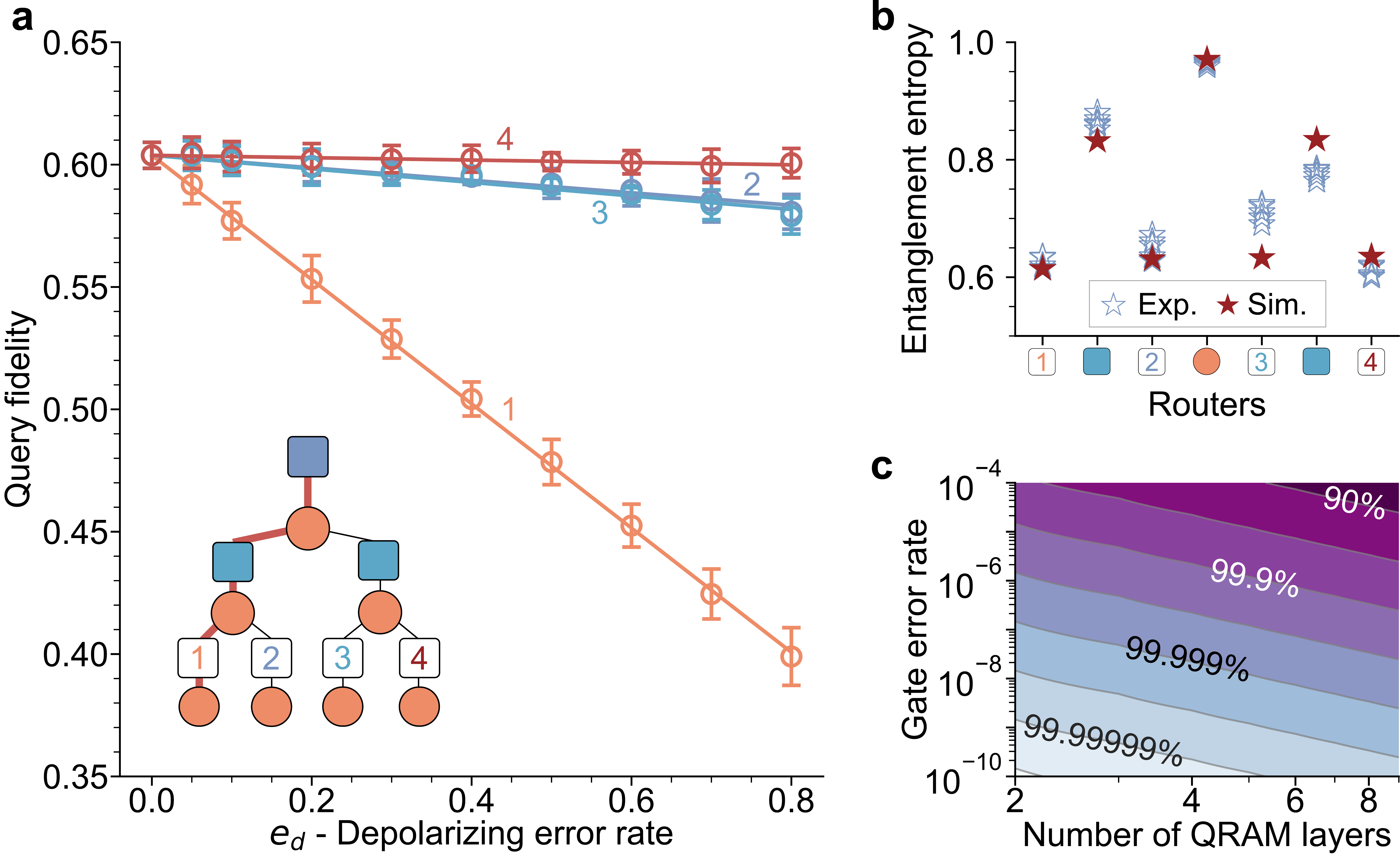}
\vspace{-.3cm}
\caption{
\textbf{Error analysis on the three-layer bucket-brigade QRAM architecture. a}, Experimental query fidelity under depolarizing errors with varying Pauli error rates.
Each time, we inject the depolarizing errors into one of the router qubits located at the third level, labeled from 1 to 4, after the data loading operation. Inset shows the schematic of the three-layer QRAM, with the queried branch highlighted.
Error bars denote standard deviation of five repeated experiments.
\textbf{b}, Von Neumann entanglement entropy of different routers in the QRAM, measured after the address loading operation.
The experimental results from five repeated runs are displayed along with the noisy simulation results.
\textbf{c}, The query fidelity as a function of the gate error rate and number of layers in the bucket-brigade architecture, obtained by noisy numerical simulation.
} 
\label{fig3}
\end{figure*}%

The router qubits in QRAM are disentangled from the address and data registers and remain in the ground state before and after a full cycle of query operation.
As a result, we can measure the router qubits after query without disturbing the quantum state in the address and data registers, and any populations detected in the router qubits signal the occurrence of error.
Based on this idea, we develop an error mitigation strategy to improve the query fidelity.
By measuring the router qubits and discarding the events where any of the router qubits is detected in the excited state, we find a significant improvement in query fidelity among all query addresses, with the average values reaching $0.800\pm0.026$, $0.782\pm0.029$ and $0.750\pm0.019$ for single-, two-, and four-component addresses (\autoref{fig2}c).
See Extended Data Fig.~\ref{edfig1} for the corresponding proportions of valid data.
In particular, by setting the query address to $\ket{\psi}_\text{A}=\frac{1}{\sqrt{2}}(\ket{00}+\ket{11})$ and configuring the classical data bits as 0101, we are able to prepare the Greenberger-Horne-Zeilinger (GHZ) state among the address and data qubits.
Applying our error mitigation approach in this scenario yields an improvement of over $0.13$ in the GHZ state fidelity, as shown in \autoref{fig2}d.

Notably, our approach provides a flexible tool for diagnosing the impact of error events that occur in different parts of the QRAM. 
By selectively applying error mitigation techniques to either the queried or unqueried branches of the QRAM architecture, we can evaluate their respective contributions to fidelity degradation.
As demonstrated in \autoref{fig2}e for the representative query addresses $\ket{0+}$ and $\ket{1+}$ (accessing left and right half of QRAM respectively), we observe a consistent fidelity improvement when error mitigation is applied to queried branches compared to unqueried ones.
The results indicate that the errors in the unqueried branches are shielded from the queried ones, reflecting the localized nature of error propagation in the bucket-brigade QRAM architecture.

\vspace{.6cm}
\noindent\textbf{\large{}Error propagation mechanism and query fidelity scaling}\\
\noindent
To gain a deeper understanding of the error propagation mechanism of the bucket-brigade QRAM, we conduct further investigation using a three-layer implementation comprising $16$ qubits (see Supplementary Information for the experimental circuit details).
We focus on querying the leftmost two among the eight classical data bits of the QRAM, which are configured to 01010101, with the query address $\ket{00+}$. 
By introducing depolarizing errors with tunable error rates to one of the four router qubits in the third layer, we quantitatively characterize the resulting degradation of the query fidelity.
Experimentally, injection of depolarizing errors is performed by randomly applying X, Y, or Z gates to the target qubit, with a probability of $p/3$ each, after the data loading phase.
The depolarizing error rate is calculated as $e_d=\frac{4p}{3}$.

The experimental results, presented in \autoref{fig3}a, reveal distinct patterns of error susceptibility across different QRAM paths.
When errors are introduced in the actively queried path, we observe a rapid, approximately linear degradation of query fidelity with increasing error rate (slope $k = -0.2549\pm0.0108$).
In contrast, error injection in the three unqueried paths demonstrates significantly more gradual fidelity decline (average slope $k=-0.0211\pm0.0069$), with the rate of degradation exhibiting a clear spatial dependence.
Remarkably, the rightmost branch, which is maximally distant from the queried path, shows nearly error-immune behavior (slope $k = -0.0053\pm0.0129$), suggesting effective error localization within the bucket-brigade architecture.
The noise resilience of the bucket-brigade architecture can be further understood from the perspective of quantum entanglement.
To this end, we measure the von Neumann entanglement entropy of the router qubits after loading the address state $\ket{+++}$ to the system.
Experimentally, the entropy is obtained as $S=-\text{Tr}[\rho \log \rho]$, where $\rho$ is the density matrix of the router qubit obtained through quantum state tomography.
As depicted in \autoref{fig3}b, the entanglement entropy decreases with increasing layer depth, reflecting the reduced quantum correlations between QRAM branches.
This reduction in entanglement confines errors to their original branches, enhancing overall noise resistance.

The inherent noise resilience of the bucket-brigade QRAM architecture makes it particularly promising for scalable implementation on noisy quantum devices.
To assess hardware requirements for scaling, we perform simulations evaluating the fidelity performance across architectures with 2 to 9 layers, while also scaling the gate error rates as shown in \autoref{fig3}c.
For efficiency, we focus on Pauli errors (see Method), with simulations conducted for maximally superposed query address state $\ket{++\cdots+}$ and a memory configuration where all classical data bits are set to 1.
\autoref{fig3}c illustrates the contour of the query fidelity in the parameter space, derived from the data in Extended Data Fig.~\ref{edfig2}.
Remarkably, the required gate error rate to maintain a target query fidelity follows a power-law decay ($\alpha\approx -2.688$) as the number of QRAM layers increases, despite the exponential growth in system qubits (see Supplementary Information for further analysis).
The results suggest that, with the future inclusion of quantum error correction techniques that is capable of exponentially suppressing errors~\cite{ai2024quantum}, the bucket-brigade QRAM could achieve near-perfect fidelity in large-scale quantum memory systems.

\vspace{.6cm}
\noindent\textbf{\large{}Conclusion and outlook}\\
\noindent
In this work, we present the experimental realization of a scalable bucket-brigade QRAM architecture on a programmable superconducting quantum processor, incorporating optimized routing operations and quantum teleportation protocols. 
By unveiling the localized nature of the error propagation mechanism and its physical origin, we provide the first experimental evidence of the inherent noise resilience in this architecture.
Furthermore, we numerically demonstrate that the bucket-brigade QRAM exhibits polylogarithmic scaling of query infidelity with respect to gate errors, which establishes the practical feasibility of implementing robust, large-scale quantum memory systems on near-term quantum hardware.

Looking ahead, as QRAM memory size grows, the exponential increase in required qubits poses challenges to current superconducting processors. 
To address this, a distributed QRAM architecture is desired.
By partitioning the QRAM across multiple processors, this approach reduces qubit and connectivity demands of a single quantum processor, bringing large-scale QRAM within reach of near-term quantum technologies. 
In particular, we note that deterministic quantum state and gate teleportation between distant superconducting chips has recently been experimentally demonstrated~\cite{Qiu2025_deterministic}.
Overall, analogous to random access memory (RAM) in classical computers, it will be necessary to design a dedicated device for QRAM in the future to overcome the exponential qubit resource.

\bibliography{ref}
\clearpage

\vspace{.6cm}
\noindent\textbf{\large{}Methods}

\vspace{2mm}
\noindent\textbf{Numerical simulation} \\
\noindent In this work, we utilize an efficient classical algorithm to numerically simulate QRAM. 
For a QRAM memory size $N$, conventional simulation methods based on the full state vector requires a complexity of $\mathcal{O}(2^N)$.
In contrast, our approach only requires a complexity of $\mathcal{O}(N_A\log N)$, where $N_\text{A}$ ($\leq N$) denotes the number of queried branches in QRAM.
The efficiency comes from the fact that the QRAM circuits are examples of efficiently computable sparse (ECS) operations~\cite{nest2010_simulating,hann2021resilience}.
Specifically, the initial state of the whole system, including the router qubits, can be expressed as a superposition of $N_A$ computational basis states:
\begin{small}
$$\sum_{i=0}^{N_\text{A}}\alpha_i\ket{j_i}_\text{A}\ket{0}_\text{D}\ket{0}_\text{R},$$ 
\end{small}
where subscripts A, D, and R denote the subsystems composed of address, data, and router qubits, respectively, and $i$ enumerates the non-zero components in the quantum address.
The QRAM circuit is composed of gates from the gate set $\mathcal{G}=\{$SWAP, $U_\text{CSWAP}$, H, X$\}$, where $U_\text{CSWAP}$ denotes the functionally equivalent CSWAP gate implemented in this work. 
These gates preserve the state of each qubit within the set $\{$$\ket{0}$, $\ket{1}$, $\ket{+}$, $\ket{-}$$\}$ for every computational basis state.
This crucial property enables efficient tracking of the system's state evolution for each address component during the entire QRAM operation.

We simulate the noisy QRAM circuit using the Monte Carlo method.
We consider Pauli error channels for each gate in $\mathcal{G}$, which keep the ECS properties of the original circuit and enable efficient noisy simulation.
For single-qubit gates H and X, we employ the depolarizing error model, where an erroneous X, Y, or Z gate is applied to the qubit after the gate with a probability of $e_s$ (also referred to as the Pauli error rate) each.
To simulate the noise of the SWAP and $U_\text{CSWAP}$ gates, we apply erroneous Pauli gates to the corresponding qubits after the gate, each of which sampled with a probability calculated based on employing the depolarizing error model with Pauli error rates of $e_s$ and $e_t$ for the single- and two-qubit gates in the corresponding decomposition circuits (see Supplementary Information for details).
In our simulation, we set $e_s=e_t/10$. The gate error rate in \autoref{fig3}c corresponds to the value of $e_t$.

\vspace{2mm}
\noindent\textbf{Quantum teleportation} \\
The embedding of an $L$-layer QRAM in a 2D square lattice requires additional $\mathcal{O}(2^L)$ ancillary qubits to be filled inbetween layers of the QRAM.
As a result, the SWAP-based communication scheme between adjacent layers can lead to a large query latency that increases the time complexity of the entire QRAM circuit.
Quantum teleportation offers an elegant way to escape the dilemma.
In our experiment, we present a proof-of-principle demonstration of the teleportation-based communication scheme in the QRAM circuit.
We use quantum teleportation to achieve the data retrieval phase of the two-layer QRAM, where the quantum information is teleported back from the source qubit ($\text{Q}_\text{s}$) to the data qubit ($\text{Q}_\text{d}$) through an ancillary qubit ($\text{Q}_\text{a}$).
Specifically, upon routing the data information into $\text{Q}_\text{s}$, we prepare $\text{Q}_\text{a}$ and $\text{Q}_\text{d}$ in the Bell state $\ket{\Phi^+}=(\ket{00}+\ket{11})/\sqrt{2}$, and then perform the Bell measurement on $\text{Q}_\text{s}$ and $\text{Q}_\text{a}$.
By post-selecting the instances where $\ket{\Phi^+}$ is measured, the quantum state is teleported from $Q_s$ to $Q_d$.
Note that the post-selection procedure can be replaced with a classical feedforward operation in the future to guarantee the scalability of the algorithm.

\vspace{.6cm}
\noindent\textbf{\large{}Data availability}\\
The data presented in the figures and that support the other findings of this study will be made publicly available for download on Zenodo/Figshare/Github upon publication.

\vspace{.5cm}
\noindent\textbf{Acknowledgement}  
{We thank Haohua Wang for supporting the device and the experimental platform on which the experiment was carried out.  The device was fabricated at the Micro-Nano Fabrication Center of Zhejiang University.
We acknowledge the support from the National Key Research and Development Program of China (Grant No. 2023YFB4502600), the Zhejiang Provincial Natural Science Foundation of China under Grant (Grant Nos. LR25F020002, LR24A040002 and LDQ23A040001), the National Natural Science Foundation of China (Grant Nos. 12174342, 12274368, 12274367, 12322414, 12404570, 12404574, 92365301). 
}

\vspace{.4cm}
\noindent\textbf{Author Contributions}\\
CS, LL, ZW, and JY conceived the project. 
FS conducted the experiments under the supervision of CS.
YJ designed the experimental circuits and performed theoretical analyses under the supervision of CS. 
YJ, FS, and DX performed simulations and analyzed the experimental data under the supervision of CS and LL.
HL fabricated the device.
CS, ZW, QG, HL, PZ, YW, KW, CZ, AZ, YZ, YG, ZC, GL, JY and YH contributed to the experimental setup. 
YJ, FS, DX, LL, and CS wrote the manuscript with input from all authors. All authors discussed the results and contributed to the final version of the manuscript.

\renewcommand{\figurename}{Extended Data Fig.}
\setcounter{figure}{0}

\begin{figure*}[p]
    \centering
    \includegraphics[width=0.83\textwidth]{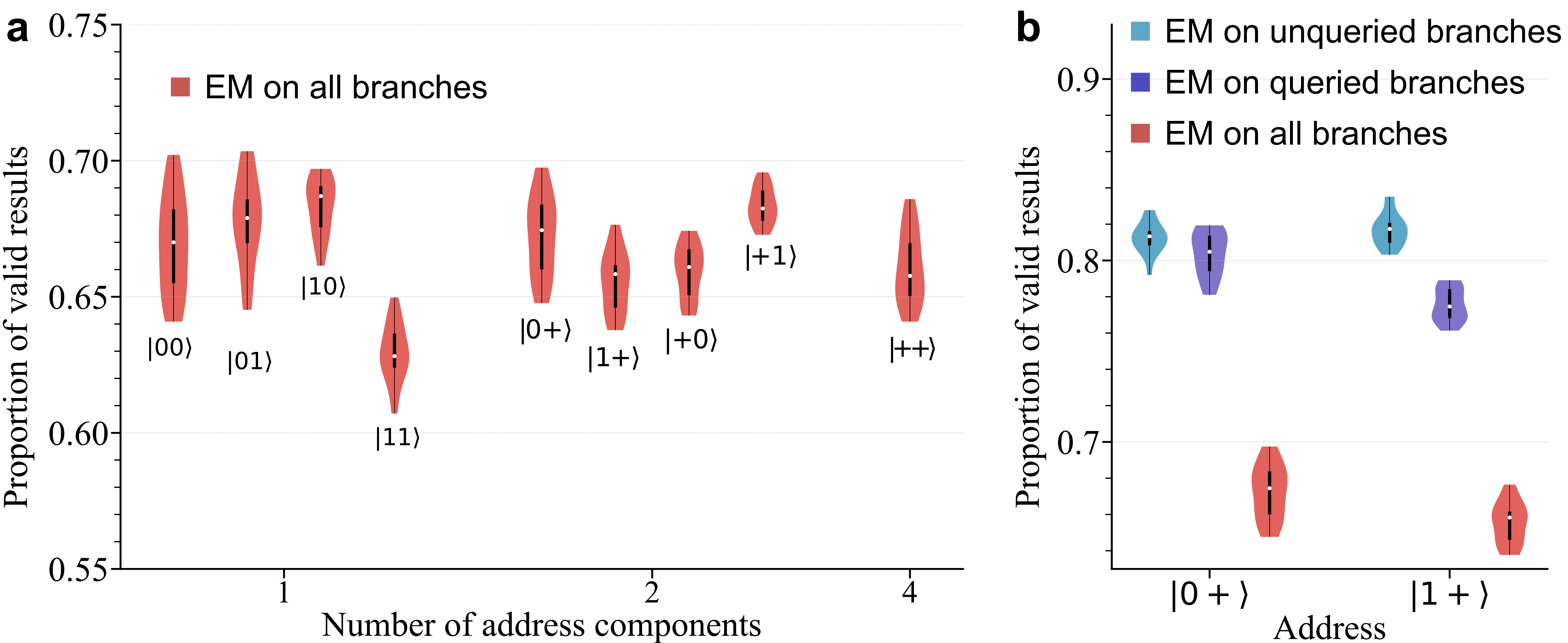}
    \vspace{-.3cm}
    \caption{\textbf{The proportion of valid results after error mitigation in the two-layer bucket-brigade QRAM.} 
    \textbf{a}, Distribution of the proportion of valid results after error mitigation for a two-layer bucket brigade QRAM under varying query addresses. \textbf{b}, Distribution of the proportion of valid results for different error mitigation strategies under query addresses $\ket{0+}$ and $\ket{1+}$.
    }
    \label{edfig1}
\end{figure*}

\clearpage

\begin{figure*}[p]
    \centering
    \includegraphics[width=0.96\textwidth]{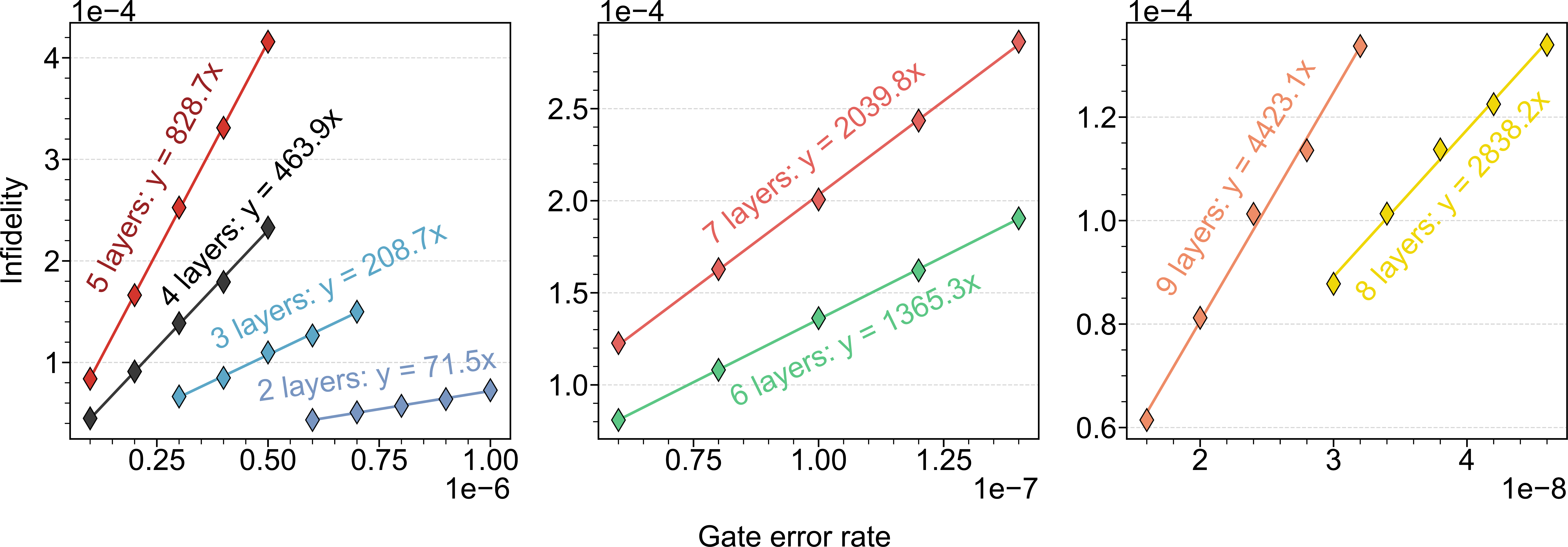}
    \caption{\textbf{The variation of the query fidelity of QRAM with respect to the error rate.} Under different levels of gate noise, the performance of the bucket-brigade QRAM is simulated and linearly fitted. Based on the linear fitting function, the gate error level required to achieve the query fidelity is calculated.
    }
    \label{edfig2}
\end{figure*}

\end{document}


\title{Supplementary Information: Experimental realization of the bucket-brigade quantum random access memory}
\maketitle
\tableofcontents

\section{Theoretical details}
\subsection{Infidelity scaling of bucket-brigade QRAM}
\label{Infidelity Scaling}
Previous work has shown that bucket-brigade QRAM is robust against timing errors, with query infidelity scaling as $\mathcal{O}(\log^3 N)$ where $N$ is the memory size~\cite{hann2021resilience}. Here we extend this analysis to prove that bucket-brigade QRAM maintains similar robustness against gate errors. Throughout this analysis, we focus on qubit-based QRAM implementations, with all logarithms being base 2.

The QRAM operation transforms an input state $\ket{\psi_{\text{in}}}$ into an output state $\ket{\psi_{\text{ideal}}}$ that encodes the requested classical data:
\begin{equation*}
    \ket{\psi_{\text{in}}} = \sum_{i=0}^{N-1} \alpha_{i} \ket{i}_{A} \ket{0}_{D} 
    \xrightarrow{\text{QRAM}} 
    \ket{\psi_{\text{ideal}}} = \sum_{i=0}^{N-1} \alpha_{i} \ket{i}_{A} \ket{x_{i}}_{D},
\end{equation*}
where $\alpha_i$ is the amplitude for address $\ket{i}$ in the superposition, $x_i$ is the classical data stored at address $\ket{i}$, and $\ket{\cdot}_A$ ($\ket{\cdot}_D$) denote address (data) qubit registers. The query fidelity is defined as
\begin{equation*}
    F = \bra{\psi_{\text{ideal}}} \text{Tr}_R(\rho) \ket{\psi_{\text{ideal}}},
\end{equation*}
where $\text{Tr}_R$ denotes the partial trace over router qubits and $\rho$ is the actual final state.

Let a single QRAM query require $T = \mathcal{O}(\log N)$ time steps. The query process, including router qubits, can be expressed as
\begin{equation*}
    \ket{\psi_{\text{ideal}}}\ket{0}_R = U_T \cdots U_2 U_1 \ket{\psi_{\text{in}}}\ket{0}_R
\end{equation*}
where $\ket{0}_R$ represents the initial router state and $U_t$ is the unitary operation at time step $t$.

Each unitary operation of QRAM can be decomposed into a gate set $\mathcal{G} = \{\text{SWAP}, \text{controlled-SWAP (CSWAP)}, H, X\}$. Additionally, qubits that aren't being operated by quantum gates stay idle. To consider the errors on these idle qubits, we apply idle gates $I$, which have a similar error rate to single-qubit gates. We model gate errors using the error channel
\begin{equation}
    \mathcal{E}(\rho) = (1-\epsilon)\rho + \sum_i K_i \rho K_i^\dagger
\end{equation}
where $K_i$ are unitary Kraus operators and $\epsilon$ is the error probability. Note that CSWAP gates require the most elementary gate operations for decomposition and thus have the highest error rates.  Without loss of generality, we assume all gates in $\mathcal{G}$ have error rates equal to that of CSWAP gates.

Let us now analyze how errors affect the quantum state during the query process. We define error configuration $c$ to specify which Kraus operator is applied to each router at each time step. The final state becomes
\begin{equation} \label{eq1}
    \rho = \sum_c p(c) \rho(c),\rho(c) = \ket{\psi(c)}\bra{\psi(c)}
\end{equation}
where $\ket{\psi(c)}$ is the final state under error configuration $c$. Thus, the query fidelity is
\begin{equation} \label{eq2}
    F = \sum_c p(c) F(c), F(c) = \bra{\psi_{\text{ideal}}} \text{Tr}_R \rho(c) \ket{\psi_{\text{ideal}}}
\end{equation}

To understand how errors propagate through the QRAM structure, we classify the queried branches. Let $\mathbf{i}$ represent the set of branches we want to query, and $\mathbf{i}(c)$ represent the subset of those branches where no direct errors occurred for a specific error configuration $c$. 
However, errors from other branches may still propagate into the good branches $\mathbf{i}(c)$. We therefore define $\tilde{\mathbf{i}}(c)$ as the maximal subset of $\mathbf{i}(c)$ where errors from other branches do not propagate into it. Under error configuration $c$, the quantum state $\ket{\psi(c)}$  can be decomposed as
\begin{equation} \label{eq3}
    \ket{\psi(c)} = \left(\sum_{i\in\tilde{\mathbf{i}}(c)} \alpha_i \ket{i}_A \ket{x_i}_D\right) \ket{f(c)}_R + \ket{\text{other}(c)}
\end{equation}.
Note that 
\begin{equation} \label{eq4}
    F(c)\geq|\Braket{\psi_{ideal},f(c)|\psi(c)}|^2,
\end{equation}
which establishes a lower bound for the fidelity $F(c)$.

We now define the "good" component of the state as
\begin{equation} \label{eq5}
    \ket{\text{good}(c)} = \left(\sum_{i\in\tilde{\mathbf{i}}(c)} \alpha_i \ket{i}_A \ket{x_i}_D\right) \ket{f(c)}_R
\end{equation}
and the corresponding overlap probabilities as
\begin{equation} \label{eq6}
    \tilde{\Lambda}(c) = \braket{\text{good}(c)|\text{good}(c)} = \sum_{i\in\tilde{\mathbf{i}}(c)} \alpha_i^2, \quad
    \Lambda(c) = \sum_{i\in\mathbf{i}(c)} \alpha_i^2
\end{equation}

Since both the applied quantum gates and noise channels are unitary, we have
\begin{equation} \label{eq7}
    |\braket{\text{other}(c)|\psi_{\text{ideal}}, f(c)}| \leq 1 - \tilde{\Lambda}(c)
\end{equation}
which leads to the fidelity lower bound
\begin{equation} \label{eq8}
    F(c) \geq (2\tilde{\Lambda}(c) - 1)^2 \quad \text{when } \tilde{\Lambda}(c) \geq \frac{1}{2}
\end{equation}
when $\tilde{\Lambda}(c)\geq \frac{1}{2}$. Next, we estimate $E(\tilde{\Lambda})$ by first estimating $E(\Lambda)$. Since we elevate all gate noise levels to the noise level of the CSWAP gate, at most $\mathcal{O}(logN)$ gates are performed on each branch at each point in time. That is to say, the probability of a single branch being error-free is greater than $(1-\epsilon)^{T\mathcal{O}(logN)}$. Through the linearity of expectation, we can derive that 
\begin{equation} \label{eq9}
    E(\Lambda)>(1-\epsilon)^{T\mathcal{O}(logN)}.
\end{equation}
Because $\tilde{\mathbf{i}}(c)\subset\mathbf{i}(c)$, we have
\begin{equation} \label{eq10}
    E(\tilde{\Lambda}) = (1-\delta)E(\Lambda)
\end{equation}
for some $\delta\in[0,1]$. Based on Eq.~\ref{eq7}--\ref{eq9}, we have
\begin{equation} \label{eq11}
    F = E(F(c)) \geq E(\sqrt{F(c)})^2\geq (2E(\tilde{\Lambda})-1)^2= (2(1-\delta)E(\Lambda)-1)^2 > (2(1-\delta)(1-\epsilon)^{Tlog(N)}-1)^2 .
\end{equation}
By applying Bernoulli inequality, we have 
\begin{equation} \label{eq12}
    1-F < 4\delta+4\epsilon T\mathcal{O}(logN),
\end{equation}
assuming that $E(\tilde{\Lambda})\geq\frac{1}{2}$. 

We can estimate $\delta$ by calculating the probability of error propagation to the good branches. Suppose a router $r$ experiences an error at time step $t$. Let $P_{r,i}(t)$ denote the probability of this error propagating to branch $i$. Then 
\begin{equation} \label{eq13}
    \delta = \epsilon\sum_{r,t}P_{r,i}(t) + \mathcal{O}(\epsilon^2).
\end{equation}
Under ideal conditions, note that the states of the qubits on the non-queried branches of the QRAM are all in the $\ket{0}$ state. This implies that when querying branch $\mathbf{i}(c)$, these errors will only propagate upwards from the left router of each router's next layer. Therefore, there are at most $\mathcal{O}(log^2N)$ such routers. When considering the worst-case scenario, where $P_{r,i}(t) = 1$, we have 
\begin{equation} \label{eq14}
    \delta\leq \epsilon T\mathcal{O}(log^2N)+\mathcal{O}(\epsilon^2)
\end{equation}
By combining Eq.~\ref{eq12} and \ref{eq14}, we obtain
\begin{equation} \label{eq15}
    1-F< 4\epsilon T(\mathcal{O}(logN) + \mathcal{O}(log^2N)).
\end{equation}
Since $T = \mathcal{O}(logN)$, we obtain that the query infidelity of QRAM based on qubits scales as $\mathcal{O}(log^3N)$. This scaling indicates that QRAM is robust even under gate noise. In particular, it is worth noting that in our proof, we elevated the error rates of all quantum gates to the level of CSWAP, and the usage of CSWAP in the query process of QRAM is also far greater than other gates. Therefore, the fidelity of the CSWAP gate is crucial for achieving high-fidelity queries.

\begin{figure}[th]
    \centering
    \includegraphics[width=0.5\linewidth]{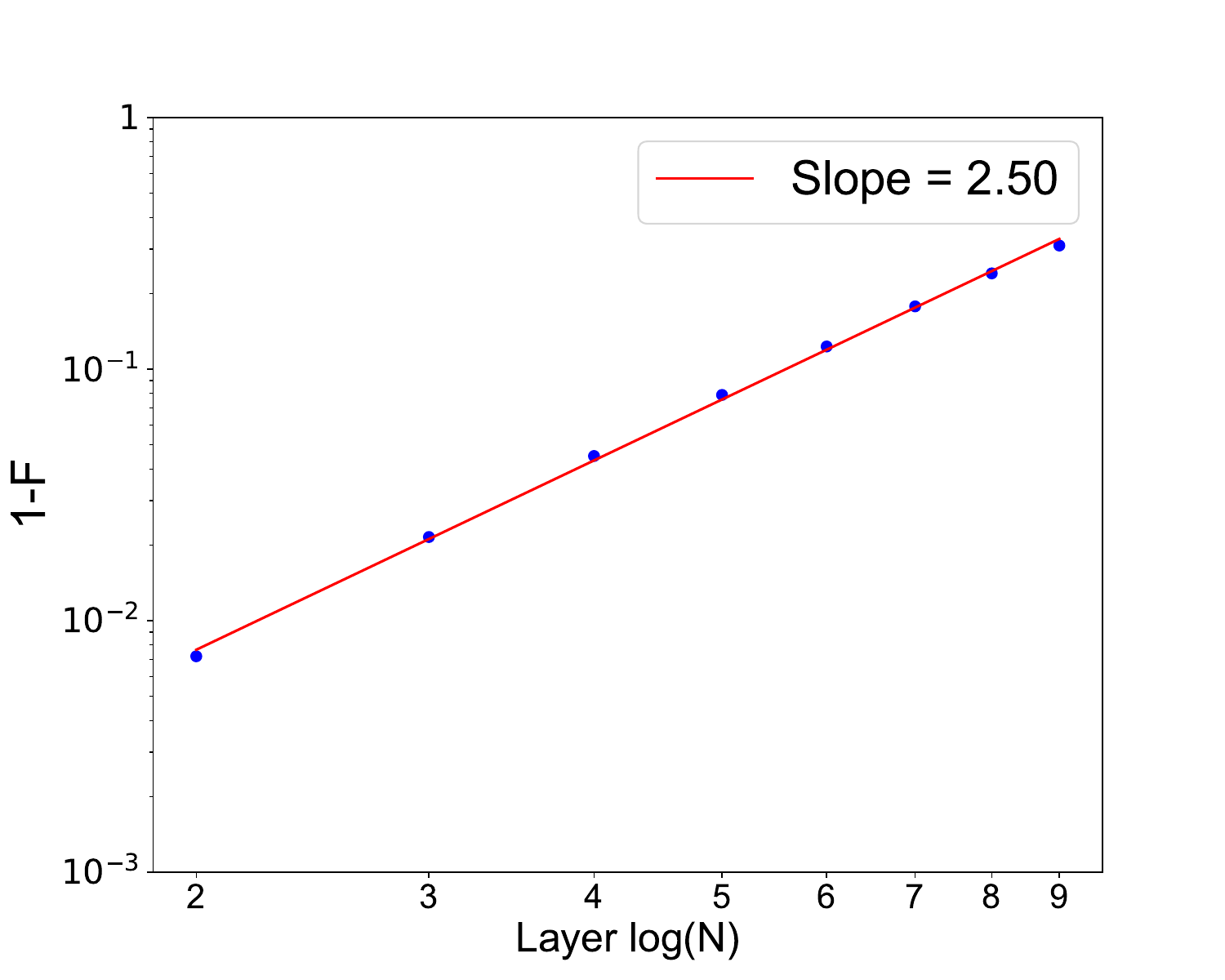}
    \caption{
    The scaling of query infidelity under gate error. The error rate $\epsilon$ of the cz gate is set to 0.0001, the classical data is $\{1,1,\cdots,1\}$, and the query address is $\ket{+\cdots+}$. Both axes are logarithmically scaled.
    }
    \label{scaling}
\end{figure}

Finally, we performed numerical simulations to verify this scaling. In the noisy simulation, noise is applied through gate errors. In our study, we considered depolarizing noise. The pauli error rate for the CZ gate is set to $\epsilon$, while the pauli error rate for single-qubit gates are set to $\frac{\epsilon}{10}$. That is,
$$\mathcal{E}_{single}(\rho) = (1-\frac{2\epsilon}{15})\rho+\frac{2\epsilon}{15}\frac{I}{2},$$
$$\mathcal{E}_{cz}(\rho) = (1-\frac{16\epsilon}{15})\rho+\frac{16\epsilon}{15}\frac{I}{4}.$$
We decomposed the SWAP and CSWAP gates into CZ and single-qubit gates, employing a specialized approach to calibrate the noise models for SWAP and CSWAP gates (see \ref{CSWAP}). In \autoref{scaling}, the slope of the fitted line is less than 3, which is consistent with the scaling. 
\vspace{-10pt}
\subsection{Quantum error mitigation}
\label{errortheory}
In the NISQ era, quantum error mitigation is an effective method to alleviate the impact of quantum noise~\cite{cai2023quantum}, which typically requires running more circuits to obtain statistical data~\cite{endo2018practical,van2023probabilistic}. Due to the unique characteristics of QRAM circuits, we propose a specially tailored quantum error mitigation technique that does not require additional circuits to evaluate noise. 

Based on Eq.~\ref{eq1}, the core idea of our error mitigation approach is to post select the cases where errors do not occur. Under ideal conditions, after the QRAM query is completed, the states of all routers should be $\ket{0}$. In the experiment of the two-layer bucket-brigade QRAM, we perform measurements on all routers after the QRAM query is completed. We select the results that yield $\ket{0}$ as valid outcomes. This approach can effectively improve the query fidelity. However, this approach is impractical for large-scale QRAM. All the noise scenarios we discuss are the same as those in \ref{Infidelity Scaling}. The probability that all routers states being $\ket{0}$ is less than $(1-\epsilon)^{T\mathcal{O}(N)} + \mathcal{O}(\epsilon)$, where $\epsilon$ is the error rate of a CSWAP gate. The exponentially decreasing probability significantly increases the number of samples, which introduces a tremendous overhead. 

We consider measuring and post-selecting only the routers in the first $\mathcal{K}$ layers. We will demonstrate that this approach is scalable and can improve query fidelity. It is noteworthy that if no errors occur in the quantum gates acting on the first two router layers, the final state of these two layers will invariably be $\ket{0}$. Hence, the probability of the router state being $\ket{0}$ in the first $\mathcal{K}$ layers is greater than $(1-\epsilon)^{T\mathcal{O}(2^\mathcal{K})}$, where $\mathcal{K}\leq log(N)$ and $\epsilon$ is the error rate of CSWAP gate. Let $\mathcal{K}_r$ represents the state of the first $\mathcal{K}$ layers of routers being $\ket{0}$. Let
\begin{equation*}
X_i = 
\begin{cases}
1 & \text{if } \;\mathcal{K}_r \\
0 &  \text{other cases}
\end{cases}
\end{equation*}
be a random variable indicating whether the result is valid. Since $p(\mathcal{K}_r) \geq (1-\epsilon)^{T\mathcal{O}(2^\mathcal{K})}$, we have $E(X_i)\geq (1-\epsilon)^{T\mathcal{O}(2^\mathcal{K})}$. Assuming we have $n$ results, let
\begin{equation*}
    \bar{X} = \frac{1}{n}\sum_i X_i.
\end{equation*}
According to the Central Limit Theorem, for a large sample size $n$, $\bar{X}$ can be approximated by a normal distribution $\mathcal{N}(p(\mathcal{K}_r), p(\mathcal{K}_r)(1-p(\mathcal{K}_r))/n)$. To achieve $|\bar{X}-p(\mathcal{K}_r)|\leq \delta$ at a 95\% confidence level, it is required that 
\begin{equation} \label{eq16}
    1.96\cdot\sqrt{Var(\bar{X})} \leq \delta \longrightarrow \; n\geq  \frac{1.96^2\cdot p(\mathcal{K}_r)(1-p(\mathcal{K}_r))}{\delta^2}.
\end{equation}
Since $p(\mathcal{K}_r) \geq (1-\epsilon)^{T\mathcal{O}(2^\mathcal{K})}$, 
\begin{equation*}
    p(\mathcal{K}_r)(1-p(\mathcal{K}_r)) \leq (1-\epsilon)^{T\mathcal{O}(2^\mathcal{K})}-(1-\epsilon)^{2T\mathcal{O}(2^\mathcal{K})}
\end{equation*}
when $p(\mathcal{K}_r)\geq\frac{1}{2}$. Assuming $\delta=0.01$, we have
\begin{equation*}
    n\geq 38416 [(1-\epsilon)^{T\mathcal{O}(2^\mathcal{K})}-(1-\epsilon)^{2T\mathcal{O}(2^\mathcal{K})}] .
\end{equation*}
In other words, even in the worst-case scenario, the number of measurements required by the above equation provides a 97.5\% probability of obtaining valid results that account for $p(\mathcal{K}_r)$ of the total. When we fix a value for $\mathcal{K}$, the proportion of valid results has a lower bound that decreases only logarithmically with $N$. Assuming that the number of valid results obtained is constant, the number of times the circuit needs to be run increases only logarithmically with $N$. This indicates that this method possesses a certain degree of scalability.

Next, we prove that this method can improve query fidelity. Equation \ref{eq1} can be rewritten as follows after applying quantum error mitigation:
\begin{equation} \label{eq17}
    \rho = \sum_c p(c|\mathcal{K}_r) \rho(c),
\end{equation}
where $p(c|\mathcal{K}_r)$ represents the conditional probability of the event given that $\mathcal{K}_r$ has occurred. Let 
\begin{equation} \label{eq18}
    \mathcal{D}_\mathcal{K} = \{c \mid \text{no errors in the first} \; \mathcal{K} \; \text{layers of routers} \}.
\end{equation}
We have
\begin{equation} \label{eq19}
    F' = \sum_cp(c|\mathcal{K}_r)F(c)\geq \sum_{c\in\mathcal{D}_\mathcal{K}}p(c|\mathcal{K}_r)F(c) = \frac{1}{p(\mathcal{K}_r)}\sum_{c\in\mathcal{D}_\mathcal{K}}p(c)F(c) = \frac{\sum_{c\in\mathcal{D}_\mathcal{K}}p(c)}{p(\mathcal{K}_r)}\sum_{c\in\mathcal{D}_\mathcal{K}}\frac{p(c)}{\sum_{c\in\mathcal{D}_\mathcal{K}}p(c)}F(c),
\end{equation}
\begin{figure}[th]
    \centering
    \includegraphics[width=1.0\textwidth]{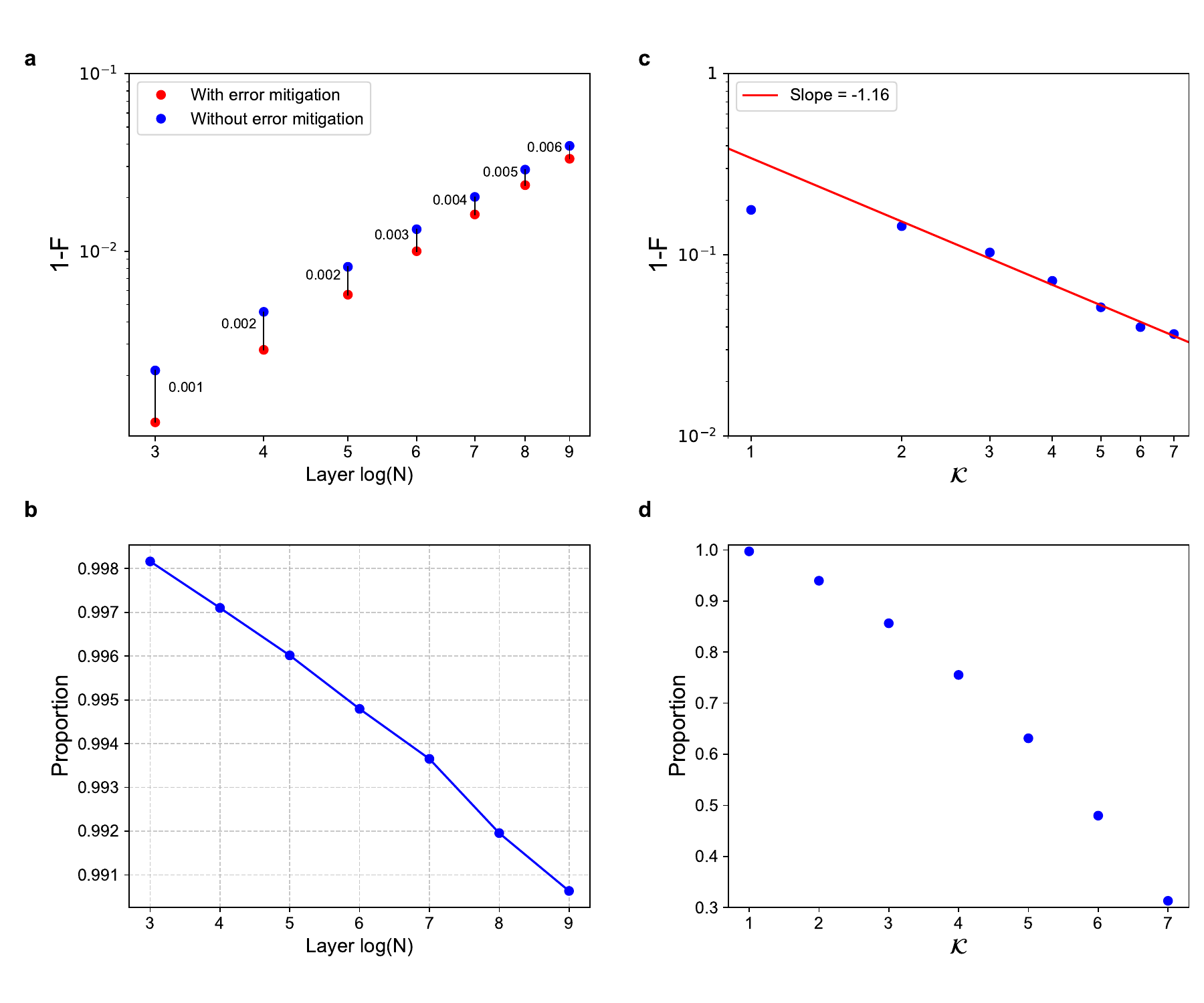}
    \caption{
    \textbf{Performance of quantum error mitigation.} All data is collected with the following conditions: classical data = $\{1,1,\cdots,1\}$ and query address = $\ket{+\cdots+}$. \textbf{a}, Comparison of query infidelity with and without the application of quantum error mitigation. Quantum error mitigation is applied to the first $\mathcal{K}=2$ layers of routers. The error rate $\epsilon$ is 0.00001, Both axes are logarithmically scaled. \textbf{b}, The proportion of valid data varies with the number of QRAM layers. Quantum error mitigation is applied to the first $\mathcal{K}=2$ layers of routers. The error rate $\epsilon$ is 0.00001. \textbf{c}, Variation of query infidelity with respect to $\mathcal{K}$. Linear fit is shown as a red line, with the corresponding slope given. Fit is performed only on data points with $\mathcal{K}\geq2$ so that the slope is not skewed by finite-size effects at small $\mathcal{K}$. Both axes are logarithmically scaled. \textbf{d}, The proportion of valid data in Subfigure \textbf{c}.
    }
    \label{error}
\end{figure}
where $F(c)=\bra{\psi_{ideal}}\rm{Tr}_R\rho(c)\ket{\psi_{ideal}}$. Note that $\sum_{c\in\mathcal{D}_\mathcal{K}}\frac{p(c)}{\sum_{c\in\mathcal{D}_\mathcal{K}}p(c)}F(c)$ represents the expectation value under the condition that no errors occur in the first $\mathcal{K}$ layers of QRAM. Referencing the proof in \ref{Infidelity Scaling}, we have
\begin{equation} \label{eq20}
    \sum_{c\in\mathcal{D}_\mathcal{K}}\frac{p(c)}{\sum_{c\in\mathcal{D}_\mathcal{K}}p(c)}F(c) \geq 1-4\epsilon T(\mathcal{O}(logN-\mathcal{K})+\mathcal{O}((logN-\mathcal{K})^2)).
\end{equation}
The probability $p(\mathcal{K}_r)$ of the router states in the first $\mathcal{K}$ layers being 0 is less than $\sum_{c\in\mathcal{D}_\mathcal{K}}p(c) + \mathcal{O}(\epsilon)$. Combining \ref{eq19} and \ref{eq20}, when $\epsilon$ is sufficiently small, we obtain 
\begin{equation} \label{eq21}
    1-F' \leq 4\epsilon T(\mathcal{O}(logN-\mathcal{K})+\mathcal{O}((logN-\mathcal{K})^2))\,, \; \mathcal{K}\leq log(N).
\end{equation}
In particular, when $\mathcal{K}=log(N)$ and $\epsilon$ is sufficiently small, $F'$ is almost equal to 1.  

The two properties mentioned above demonstrate that our quantum error mitigation approach is effective and exhibits a certain degree of scalability. To validate our conclusions, we conducted numerical simulations. In \autoref{error}a, we applied quantum error mitigation to the first two layers of routers. It can be observed that when $\mathcal{K}$ is fixed, as the scale of QRAM increases, the effect of quantum error mitigation gradually deteriorates, which is consistent with Eq.~\ref{eq21}. In \autoref{error}b, we perform quantum error mitigation with different $\mathcal{K}$ for a QRAM with $N=128$. It can be observed that when $N$ is fixed, as $\mathcal{K}$ increases, the effect of quantum error mitigation gradually improves. Moreover, under logarithmic scaling, it exhibits a linear trend with a slope greater than -2, which is also consistent with Eq.~\ref{eq21}. It is worth noting that when $\mathcal{K}=7$, the fidelity after applying quantum error mitigation does not reach 1. This is because when the error rate of the two single-qubit gates is 0.0001, the error rate of the CSWAP gate is approximately 10 times that of the two-qubit gate error rate, which is not sufficiently small for us to ignore the influence of $\mathcal{O}(\epsilon)$ in $p(\mathcal{K}_r)$. In \autoref{error}c, we plot the proportion of valid data as it decreases with increasing $\mathcal{K}$. The rate of decrease accelerates as $\mathcal{K}$ grows larger, which aligns with our expectations. In \autoref{error}d, we can observe that as the number of QRAM layers increases, although the number of qubits grows exponentially, the proportion of valid data only decreases slightly when $\mathcal{K}$ is fixed. When $\epsilon$ is smaller, this decline becomes even less noticeable.

\subsection{Distributed QRAM}
\begin{figure*}[th]
\centering
\includegraphics[width=0.75\textwidth]{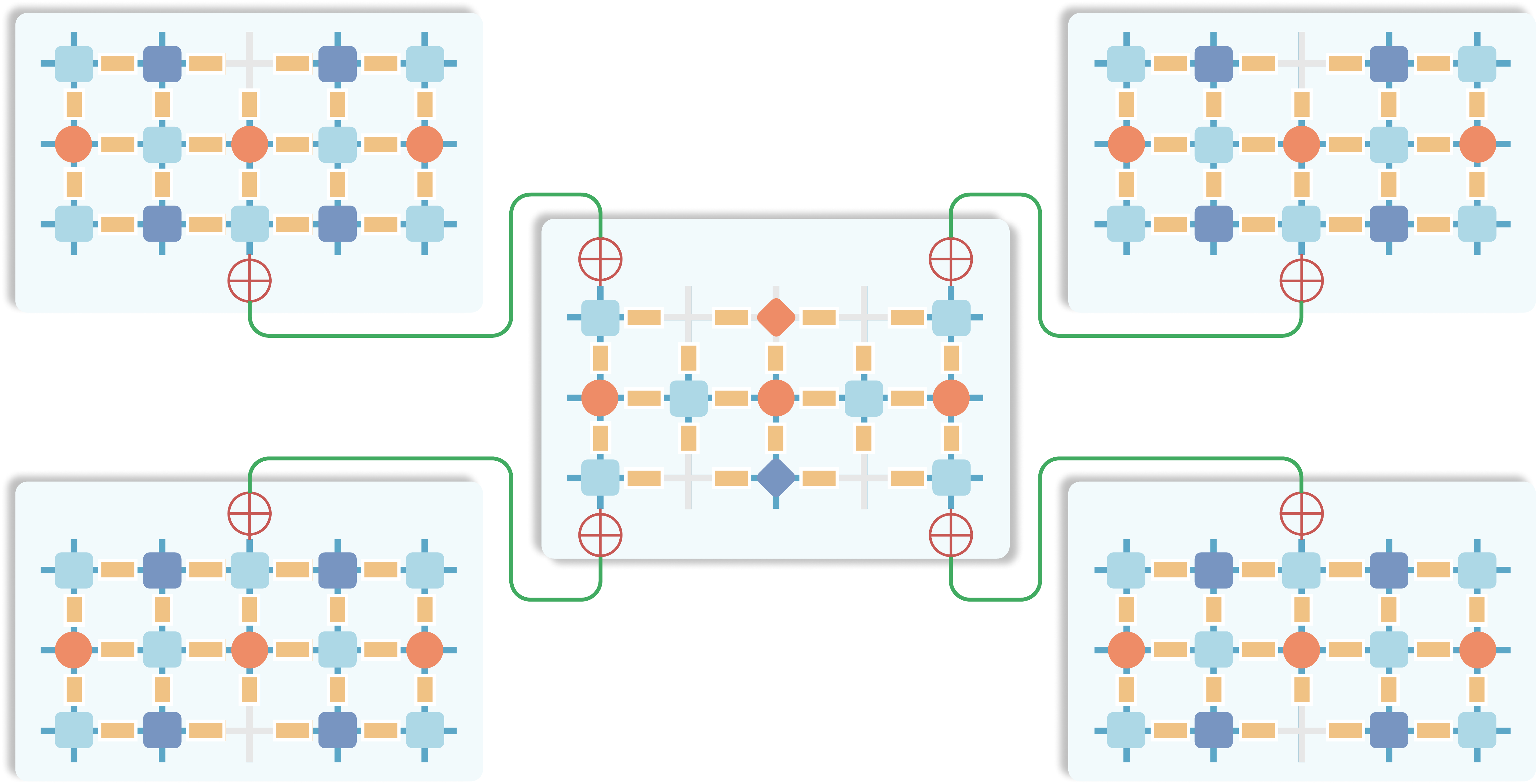}
\caption{
\textbf{ 
Four-layer distributed QRAM implemented via remote CZ gate. }Yellow rectangles with crosses: qubit SQUID (with asymmetric
junction sizes); red circles: tunable couplers; crosses inside: Josephson junctions; green lines: superconducting cables; light blue squares: Router Data; orange circles: Router Address; dark blue diamonds: qubits in data registers; orange squares: qubits in address registers; dark blue squares: qubits in the last QRAM layer for reading classical data. The original structure of the QRAM consists of routers each with one Router Address and two Router Data, as described in~\cite{hann2019hardware}.}
\label{Distributed}
\end{figure*}

The natural structure of QRAM is a binary tree~\cite{giovannetti2008quantum}. However, on real quantum devices, there is no native topological structure that can naturally implement QRAM. Considering the topological connectivity of qubits, the H-tree recursive mapping necessitates the introduction of a substantial number of auxiliary qubits for QRAM implementation. Therefore, we propose a distributed QRAM scheme that enables efficient QRAM realization while introducing minimal auxiliary qubits. As illustrated in \autoref{Distributed}, we implement small sub-trees of the QRAM on different superconducting quantum chips. When routers located on two separate chips need to interact, we prepare two qubits in a Bell state. This Bell state enables the transfer of a CZ gate between the two routers, effectively realizing a remote CZ gate~\cite{PhysRevLett.93.240501,doi:10.1126/science.aaw9415}. The interconnection between different superconducting quantum chips and the implementation of remote quantum gates have already been realized~\cite{QIU2025351,Niu2023}. This distributed QRAM architecture facilitates a modular design. Furthermore, it necessitates the addition of only two auxiliary qubits specifically at the points where remote connections between two chips are required, thereby significantly reducing the overall number of auxiliary qubits.

\section{Experimental details}

\subsection{Device Information}
\begin{figure*}[th]
\centering
\includegraphics[width=0.83\textwidth]{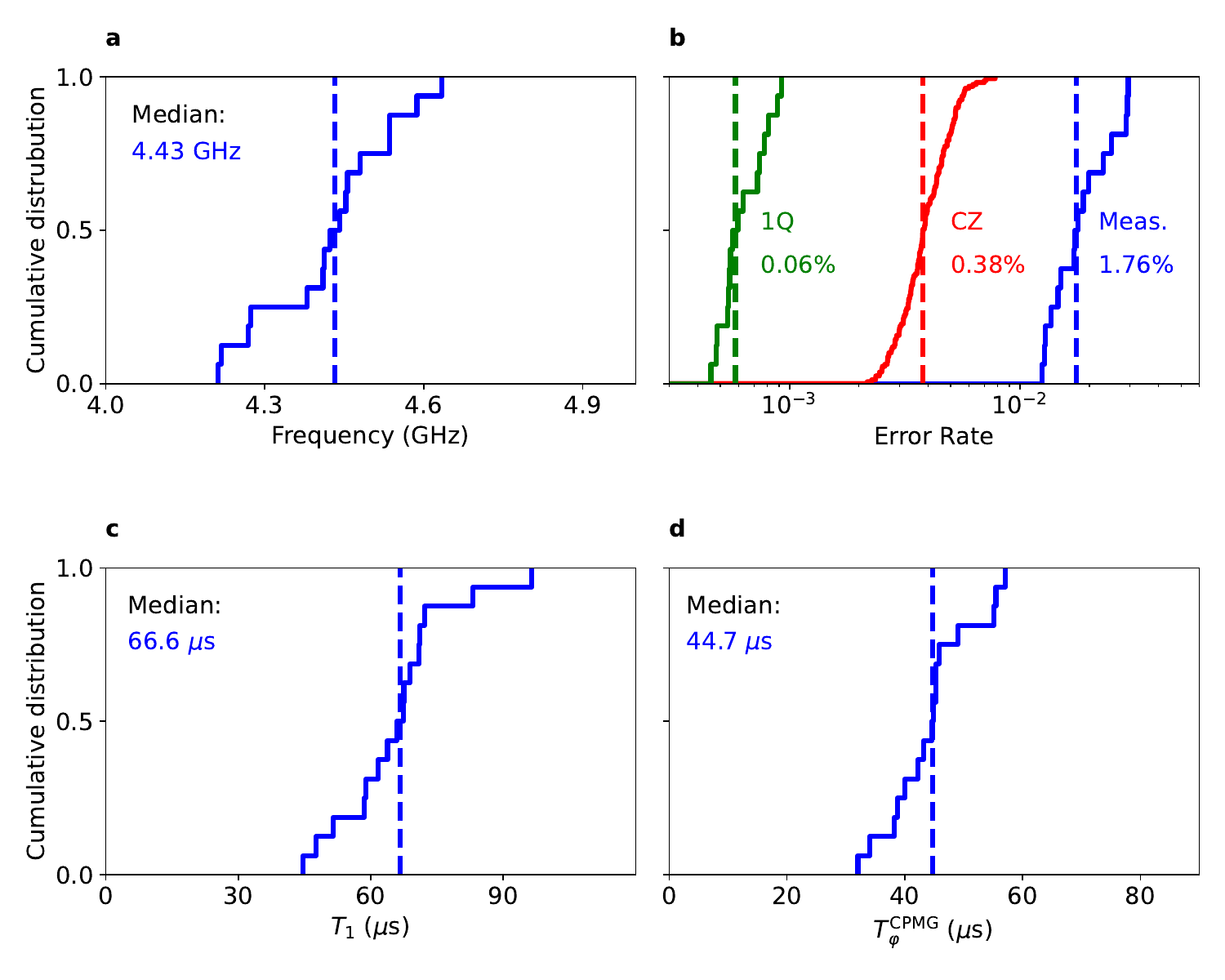}
\vspace{-.8cm}
\caption{
\textbf{ 
Cumulative distribution of qubit parameters. a,} Qubit idle frequencies.
\textbf{
b,} Pauli errors of the simultaneous single-qubit gate, Pauli errors of the CZ gate, and simultaneous average measurement identification errors.
\textbf{
c,} Qubit relaxation time measured at the corresponding idle frequencies.
\textbf{
d,} Qubit dephasing time measured using CPMG sequence at the corresponding idle frequencies.
}
\label{fig_supp_device_param}
\end{figure*}

Our experiments are conducted on a superconducting quantum processor that shares the same layout as the processor described in ~\cite{jin2025observationtopologicalprethermalstrong}, but features an enhanced qubit-resonator coupling strength, enabling readout time of about 500 ns. The inclusion of the Purcell filter ensures high qubit coherence is maintained, despite the increased coupling strength and bandwidth of the readout resonators. Each qubit is equipped with an individual microwave control line for executing single-qubit gates and a flux control line for frequency modulation. Furthermore, each coupler is independently controlled via a dedicated flux line, allowing precise adjustment of the coupling strength between neighboring qubits. As illustrated in \autoref{struct}b, we employ 16 qubits to implement two- and three-level QRAM in our experiments. Detailed properties of these qubits, including their gate errors and simultaneous average measurement identification errors, are presented in \autoref{fig_supp_device_param}.

\subsection{Routing Operation}
\label{CSWAP}

The CSWAP gate serves as the fundamental building block for all routing operations in bucket-brigade QRAM architectures~\cite{chen2023efficient}. However, synthesizing CSWAP gates from CZ gates and single-qubit rotations incurs substantial overhead~\cite{chau1995simple,mq2024shallow,saha2023intermediate}. Without considering topological constraints, the CSWAP gate requires a minimum of 7 CZ gates for implementation~\cite{chau1995simple}. When accounting for qubit connectivity on a square lattice topology, this increases to 8 CZ gates~\cite{mq2024shallow}. In practice, two-qubit CZ gates typically exhibit lower fidelity and require more complex control mechanisms compared to single-qubit operations. The extensive use of CSWAP gates in QRAM poses a significant challenge for achieving high query fidelity.

\begin{figure*}[th]
    \centering
    \includegraphics[width=0.5\textwidth]{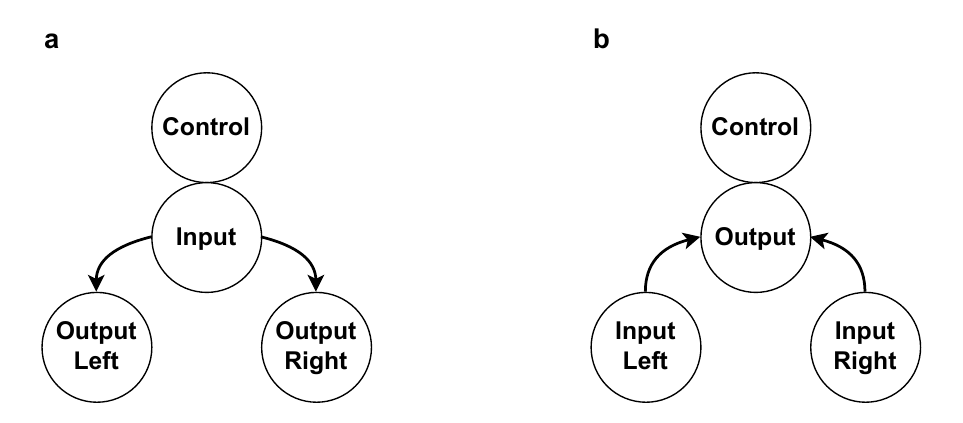}
    \caption{\textbf{The two distinct routing operations in QRAM.} Both operations consist of two CSWAP gates, where the routing is directed to the left when the control qubit is $\ket{0}$ and to the right when it is $\ket{1}$. The target qubits serve as the Input and Output for these operations. \textbf{a}, Downward routing. The valid data propagates downwards. \textbf{b}, Upward routing. The valid data propagates upwards.
    }
    \label{Routing}
\end{figure*}

Fortunately, due to the unique circuit structure of QRAM, implementing a complete CSWAP operation is not necessary, which enables substantial optimization. We identify two distinct routing scenarios: downward routing and upward routing, as illustrated in \autoref{Routing}. In downward routing, the output qubit is always in state $\ket{0}$ before the routing operation. For upward routing, within the computational subspace where the control qubit is either $\ket{0}$ or $\ket{1}$, the output qubit consistently remains in the $\ket{0}$ state before the routing operation. This constraint ensures that within the active subspace (where the control qubit is $\ket{1}$), one target qubit is always guaranteed to be $\ket{0}$.

Building on this observation, we first replace the standard CSWAP gate with a unitary operation that satisfies the following constraints:
\begin{equation*}
\begin{aligned}
        U'\ket{0}_{\text{control}}\ket{\psi}_{\text{target}} &= \ket{0}_{\text{control}}\ket{\psi}_{\text{target}}\\
        U'\ket{1}_{\text{control}}\ket{00}_{\text{target}} &= \ket{1}_{\text{control}}\ket{00}_{\text{target}}\\
        U'\ket{1}_{\text{control}}\ket{10}_{\text{target}} &= \ket{1}_{\text{control}}\ket{01}_{\text{target}},
\end{aligned}
\end{equation*}
That is to say, $U'$ possesses two degrees of freedom on the computation bases $\ket{101}$ and $\ket{111}$. When we restrict $U'$ to the subspace constructed by these two bases, and noting that $U'$ is a unitary matrix, we have
\begin{equation*}
\begin{aligned}
    U'_{|span\{\ket{101},\ket{111}\}}: span\{\ket{101},\ket{111}\}&\rightarrow span\{\ket{110},\ket{111}\},\\
    [U'_{|span\{\ket{101},\ket{111}\}}]_{span\{\ket{110},\ket{111}\}}^{span\{\ket{101},\ket{111}\}} &= \begin{bmatrix}
        a & b \\
        -e^{i\theta}b^* & e^{i\theta} a^*
    \end{bmatrix},
\end{aligned}
\end{equation*}
where $a,b\in \mathcal{C}$ and $a^2+b^2=1$.
We utilized Cpflow~\cite{nemkov2023efficient} to decompose unitary matrices that satisfy the conditions and find a unitary matrix that minimizes the number of required CZ gates
\[
\renewcommand{\arraystretch}{0.85}
    U'=\begin{bmatrix}
        1 & 0 & 0 & 0 & 0 & 0 & 0 & 0 \\
        0 & 1 & 0 & 0 & 0 & 0 & 0 & 0 \\
        0 & 0 & 1 & 0 & 0 & 0 & 0 & 0 \\
        0 & 0 & 0 & 1 & 0 & 0 & 0 & 0 \\
        0 & 0 & 0 & 0 & 1 & 0 & 0 & 0 \\
        0 & 0 & 0 & 0 & 0 & 0 & 1 & 0 \\
        0 & 0 & 0 & 0 & 0 & 1 & 0 & 0 \\
        0 & 0 & 0 & 0 & 0 & 0 & 0 & -1
    \end{bmatrix}.
\]
The decomposition of this $U'$ only requires 5 CZ gates when the control qubit connects to one target qubit, and 8 CZ gates when connected to two target qubits (all decompositions available online in QASM file format). In contrast, the standard CSWAP gate requires 8 and 10 CZ gates under the same topological constraints, respectively~\cite{mq2024shallow}.

\begin{table}[h!]
\centering
\renewcommand{\arraystretch}{1.2} 
\setlength{\tabcolsep}{4pt} 
\begin{tabular}{|c|c|c|c|}
\hline
\textbf{Operator} & \textbf{Connectivity} & \textbf{CZ count} & \textbf{Fidelity} \\ \hline
\begin{minipage}[c][1.2cm][c]{3cm}
\centering
$U'$
\end{minipage} & 
\begin{minipage}[c][1.2cm][c]{4cm}
\centering
  \scalebox{0.8}{
\begin{tikzpicture}[node distance=1.4cm]
    \node[draw, circle] (A) {Control};
    \node[draw, circle, right of=A] (B) {Target};
    \node[draw, circle, right of=B] (C) {Control};
    
    \draw (A) -- (B);
    \draw (B) -- (C);
\end{tikzpicture}
}
\end{minipage} & 
\begin{minipage}[c][1.2cm][c]{1.8cm}
\centering
5
\end{minipage} & 
\begin{minipage}[c][1.2cm][c]{1.8cm}
\centering
0.9547
\end{minipage} \\ \hline

\begin{minipage}[c][1.2cm][c]{3cm}
\centering
$U'$
\end{minipage} & 
\begin{minipage}[c][1.2cm][c]{4cm}
\centering
  \scalebox{0.8}{
\begin{tikzpicture}[node distance=1.4cm]
    \node[draw, circle] (A) {Target};
    \node[draw, circle, right of=A] (B) {Control};
    \node[draw, circle, right of=B] (C) {Target};
    
    \draw (A) -- (B);
    \draw (B) -- (C);
\end{tikzpicture}
}
\end{minipage}
& 
\begin{minipage}[c][1.2cm][c]{1.8cm}
\centering
8
\end{minipage} & 
\begin{minipage}[c][1.2cm][c]{1.8cm}
\centering
0.9330
\end{minipage} \\ \hline

\begin{minipage}[c][1.2cm][c]{3cm}
\centering
$U''$
\end{minipage} & 
\begin{minipage}[c][1.2cm][c]{4cm}
\centering
  \scalebox{0.8}{
\begin{tikzpicture}[node distance=1.4cm]
    \node[draw, circle] (A) {Target};
    \node[draw, circle, right of=A] (B) {Control};
    \node[draw, circle, right of=B] (C) {Target};
    
    \draw (A) -- (B);
    \draw (B) -- (C);
\end{tikzpicture}
}
\end{minipage}
& 
\begin{minipage}[c][1.2cm][c]{1.8cm}
\centering
7
\end{minipage} & 
\begin{minipage}[c][1.2cm][c]{1.8cm}
\centering
0.9384
\end{minipage} \\ \hline

\begin{minipage}[c][1.2cm][c]{3cm}
\centering
Fredkin
\end{minipage} & 
\begin{minipage}[c][1.2cm][c]{4cm}
\centering
  \scalebox{0.8}{
\begin{tikzpicture}[node distance=1.4cm]
    \node[draw, circle] (A) {Control};
    \node[draw, circle, right of=A] (B) {Target};
    \node[draw, circle, right of=B] (C) {Target};
    
    \draw (A) -- (B);
    \draw (B) -- (C);
\end{tikzpicture}
}
\end{minipage}
& 
\begin{minipage}[c][1.2cm][c]{1.8cm}
\centering
8
\end{minipage} & 
\begin{minipage}[c][1.2cm][c]{1.8cm}
\centering
0.9324
\end{minipage} \\ \hline

\begin{minipage}[c][1.2cm][c]{3cm}
\centering
Fredkin
\end{minipage} & 
\begin{minipage}[c][1.2cm][c]{4cm}
\centering
  \scalebox{0.8}{
\begin{tikzpicture}[node distance=1.4cm]
    \node[draw, circle] (A) {Target};
    \node[draw, circle, right of=A] (B) {Control};
    \node[draw, circle, right of=B] (C) {Target};
    
    \draw (A) -- (B);
    \draw (B) -- (C);
\end{tikzpicture}
}
\end{minipage} & 
\begin{minipage}[c][1.2cm][c]{1.8cm}
\centering
10
\end{minipage} & 
\begin{minipage}[c][1.2cm][c]{1.8cm}
\centering
0.9187
\end{minipage} \\ \hline

\begin{minipage}[c][2.2cm][c]{3cm}
\centering
Routing with $U'$
\end{minipage} & 
\begin{minipage}[c][2.2cm][c]{4cm}
\centering
  \scalebox{0.8}{
\begin{tikzpicture}[node distance=1.4cm]
    \node[draw=red, circle] (Center) {Target};
    \node[draw, circle, above of=Center] (Top) {Control};
    \node[draw, circle, left of=Center] (Left) {Target};
    \node[draw, circle, right of=Center] (Right) {Target};
    
    \draw (Center) -- (Top);
    \draw (Center) -- (Left);
    \draw (Center) -- (Right);
\end{tikzpicture}
}
\end{minipage} & 
\begin{minipage}[c][1.8cm][c]{1.8cm}
\centering
10
\end{minipage} & 
\begin{minipage}[c][1.8cm][c]{1.8cm}
\centering
0.8935
\end{minipage} \\ \hline
\begin{minipage}[c][1.8cm][c]{3cm}
\centering
Routing with Fredkin
\end{minipage} & 
\begin{minipage}[c][2.2cm][c]{4cm}
\centering
  \scalebox{0.8}{
\begin{tikzpicture}[node distance=1.4cm]
    \node[draw=red, circle] (Center) {Target};
    \node[draw, circle, above of=Center] (Top) {Control};
    \node[draw, circle, left of=Center] (Left) {Target};
    \node[draw, circle, right of=Center] (Right) {Target};
    
    \draw (Center) -- (Top);
    \draw (Center) -- (Left);
    \draw (Center) -- (Right);
\end{tikzpicture}
}
\end{minipage} & 
\begin{minipage}[c][1.8cm][c]{1.8cm}
\centering
16
\end{minipage} & 
\begin{minipage}[c][1.8cm][c]{1.8cm}
\centering
0.8529
\end{minipage} \\ \hline

\begin{minipage}[c][1.8cm][c]{3cm}
\centering
Routing with $U'''$
\end{minipage} & 
\begin{minipage}[c][2.2cm][c]{4cm}
\centering
  \scalebox{0.8}{
\begin{tikzpicture}[node distance=1.4cm]
    \node[draw, circle] (Center) {Control};
    \node[draw=red, circle, above of=Center] (Top) {Target};
    \node[draw, circle, left of=Center] (Left) {Target};
    \node[draw, circle, right of=Center] (Right) {Target};
    
    \draw (Center) -- (Top);
    \draw (Center) -- (Left);
    \draw (Center) -- (Right);
\end{tikzpicture}
}
\end{minipage} & 
\begin{minipage}[c][1.8cm][c]{1.8cm}
\centering
12
\end{minipage} & 
\begin{minipage}[c][1.8cm][c]{1.8cm}
\centering
0.8848
\end{minipage} \\ \hline

\begin{minipage}[c][1.8cm][c]{3cm}
\centering
Routing with Fredkin
\end{minipage} & 
\begin{minipage}[c][2.2cm][c]{4cm}
\centering
  \scalebox{0.8}{
\begin{tikzpicture}[node distance=1.4cm]
    \node[draw, circle] (Center) {Control};
    \node[draw=red, circle, above of=Center] (Top) {Target};
    \node[draw, circle, left of=Center] (Left) {Target};
    \node[draw, circle, right of=Center] (Right) {Target};
    
    \draw (Center) -- (Top);
    \draw (Center) -- (Left);
    \draw (Center) -- (Right);
\end{tikzpicture}
}
\end{minipage} & 
\begin{minipage}[c][1.8cm][c]{1.8cm}
\centering
20
\end{minipage} & 
\begin{minipage}[c][1.8cm][c]{1.8cm}
\centering
0.8218
\end{minipage} \\ \hline

\end{tabular}
\caption{\textbf{Number of CZ gates and fidelity for different operations under various topological connections.} In the "Connectivity" column of the routing operation, a red circle indicates that the corresponding qubit is acted upon twice.}
\label{CSWAPFidelity}
\end{table}

Furthermore, we can optimize the upward routing and downward routing separately using this  , rather than seeking a universal routing operation. For downward routing, since one target qubit is necessarily $\ket{0}$, any unitary matrix whose action on the basis states $\{\ket{000},\ket{010},\ket{100},\ket{110}\}$ matches the CSWAP gate can be employed. Following the same optimization procedure and under the topology where one control qubit connects to two target qubits, we identified an optimal matrix
\[
\renewcommand{\arraystretch}{0.85}
    U''=\begin{bmatrix}
        1 & 0 & 0 & 0 & 0 & 0 & 0 & 0 \\
        0 & 0 & 0 & 0 & 0 & 0 & 0 & -1 \\
        0 & 0 & 1 & 0 & 0 & 0 & 0 & 0 \\
        0 & 0 & 0 & 0 & 0 & 1 & 0 & 0 \\
        0 & 0 & 0 & 0 & 1 & 0 & 0 & 0 \\
        0 & 0 & 0 & 0 & 0 & 0 & 1 & 0 \\
        0 & 0 & 0 & 1 & 0 & 0 & 0 & 0 \\
        0 & 1 & 0 & 0 & 0 & 0 & 0 & 0
    \end{bmatrix}
\]
It requires only 7 CZ gates to implement the downward routing operation. Additionally, it can also be employed for upward routing when the output qubit is not occupied by data from another branch. To further reduce the number of CZ gates, we optimized the upward routing for four qubits. Based on the observation that the output qubit is $\ket{0}$, and following a similar procedure as above, we have identified a unitary operation $U'''$ that requires only 12 CZ gates when the control qubit is connected to the remaining three qubits.

\begin{figure*}[th]
    \centering
    \includegraphics[width=1\textwidth]{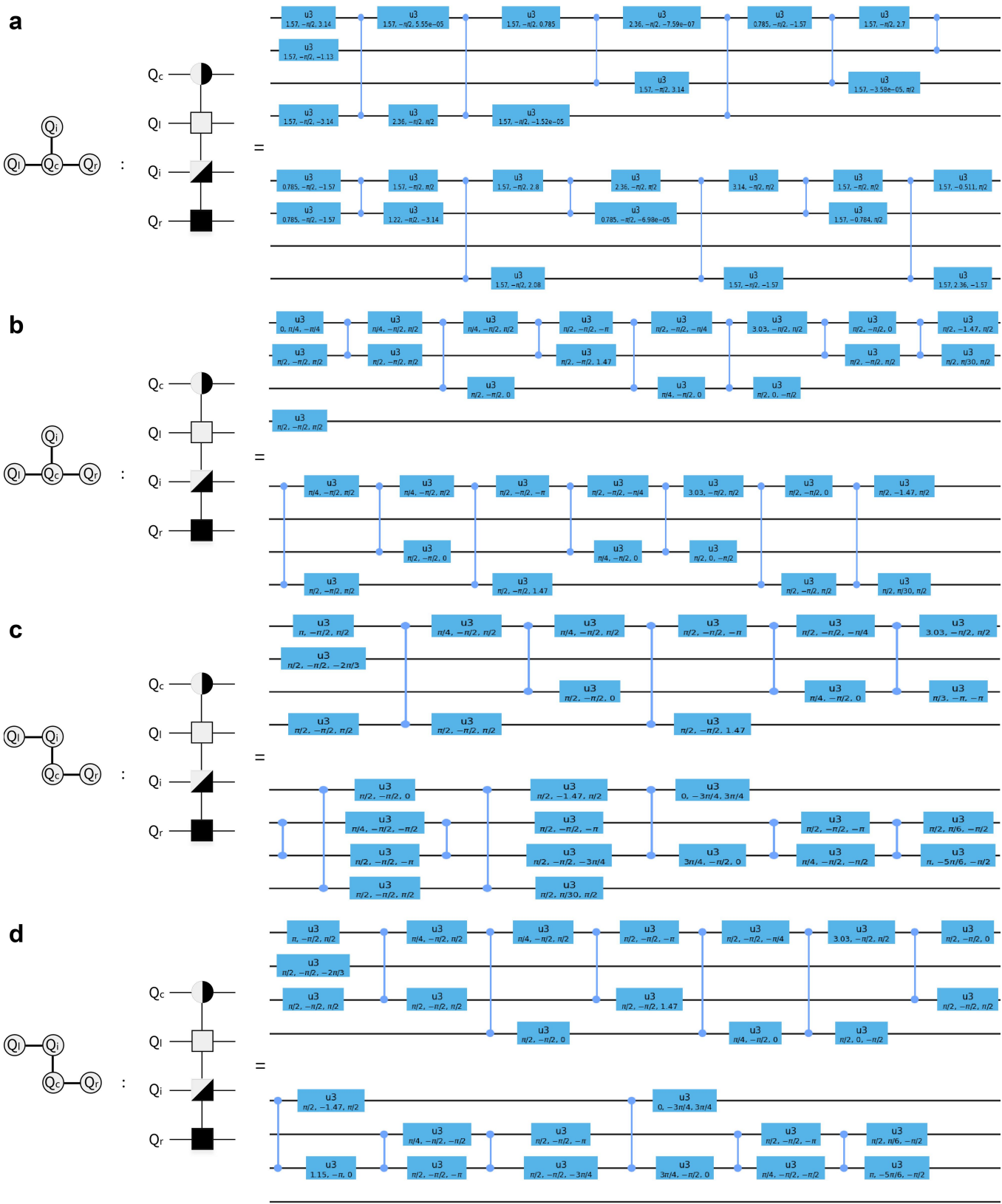}
    \caption{\textbf{Circuit decomposition for routing operations under various topological connections and application scenarios.} \textbf{a}, Routing operation for upword routing. \textbf{b}, Routing operation for downword routing. \textbf{c}, Routing operation for downword routing. \textbf{d}, Routing operation for upword routing.
    }
    \label{routing under topology}
\end{figure*}

Next, we experimentally performed fidelity estimation for $U'$ synthesized based on CZ and single-qubit gates. We evaluate the fidelity of synthesizing $U'$ and CSWAP through CZ gates and single-qubit gates by employing the quantum process tomography numerically. It can be observed that the fidelity of routing operations can be improved through the utilization of $U'$. In \autoref{CSWAPFidelity}, we can observe that under the same topological connections, the fidelity of $U'$ is significantly higher than that of CSWAP. Due to the large number of routing operations in the QRAM query process, the advantage of $U'$ will be further amplified. Using these equivalent matrices, we achieved routing operations under several topological connections, as shown in \autoref{routing under topology}.

Finally, we will elucidate how we apply the noise of $U'$ in large-scale QRAM numerical simulations. As mentioned in the Methods section of the main text under Numerical simulation, our noisy simulations are limited to the case where $$K\ket{i}=\ket{i'},$$ with $K$ denoting the Kraus operator and $\ket{i}$, $\ket{i'}\in\{$$\ket{0}$, $\ket{1}$, $\ket{+}$, $\ket{-}$$\}$. The single-qubit gates and CZ gates employ depolarizing noise, and the noise channels of each gate are not necessarily commutative with the subsequent gate. This implies that it may not be possible to derive a noise channel for the $U'$ that satisfies the simulation conditions. A simple idea is to calculate the probability of all gates operating without error as the probability of the $U'$ functioning correctly, and then apply 3-qubit depolarizing noise at the end of the $U'$. However, this approach lacks precision and may not accurately capture the noise characteristics of the $U'$. As shown in Fig.1b of the main text, the number of quantum gates applied to the control qubits is significantly fewer than those applied to the target qubit; therefore, directly applying uniform depolarizing noise is inaccurate. We adopt an approach where all the noise from the gates used in synthesizing $U'$ is applied after the completion of $U'$. In \autoref{Comparation of noise}, we prepare the density matrices of the ideal noisy circuits under different initial states and compare them with the density matrices obtained by applying the two aforementioned noise application methods after the ideal $U'$. We observe that sequentially applying gate noise after the application of $U'$ yields more accurate results.
\begin{center}
\begin{table}[h]
\begin{tabular}{|c|c|c|}
\hline
\textbf{Initial state} & \textbf{Fidelity under uniform depolarizing noise} & \textbf{Fidelity under sequential application of gate noise} \\ \hline
$\ket{000}$ & 0.98589 & 0.99659 \\
$\ket{111}$ & 0.98589 & 0.99659 \\
$\ket{+++}$ & 0.99723 & 0.99723 \\
$\ket{---}$ & 0.99723 & 0.99723 \\
$\ket{+i}\ket{+i}\ket{+i}$ & 0.99726 & 0.99732 \\
$(\ket{000}+\ket{111})/\sqrt{2}$ & 0.99640 & 0.99775 \\
\hline
\end{tabular}
\caption{\textbf{A comparative study of two noise application methods.} We conduct a comparative analysis under the conditions of 99.2\% fidelity for the CZ gate and 99.92\% fidelity for the single-qubit gate. In the initial state, $\ket{+}=(\ket{0}+\ket{1})/\sqrt{2}$, $\ket{-}=(\ket{0}-\ket{1})/\sqrt{2}$ and $\ket{+i}=(\ket{0}+i\ket{1})/\sqrt{2}$.}
\label{Comparation of noise}
\end{table}
\end{center}

\vspace{-16pt}
\subsection{Readout error calibration}
In our experimental implementation, readout error correction is performed using calibration matrices. The single-qubit response matrix $R_{ij} = P(m=i|t=j)$ was constructed from measured readout fidelities, where $P(m=i|t=j)$ represents the conditional probability of measuring state $\ket{i}$ given the prepared state $\ket{j}$. The measured probability distributions are then corrected by applying the inverse of this response matrix.

\subsection{Experimental layout of qubits in QRAM}
\begin{figure*}[th]
    \centering
    \includegraphics[width=0.98\textwidth]{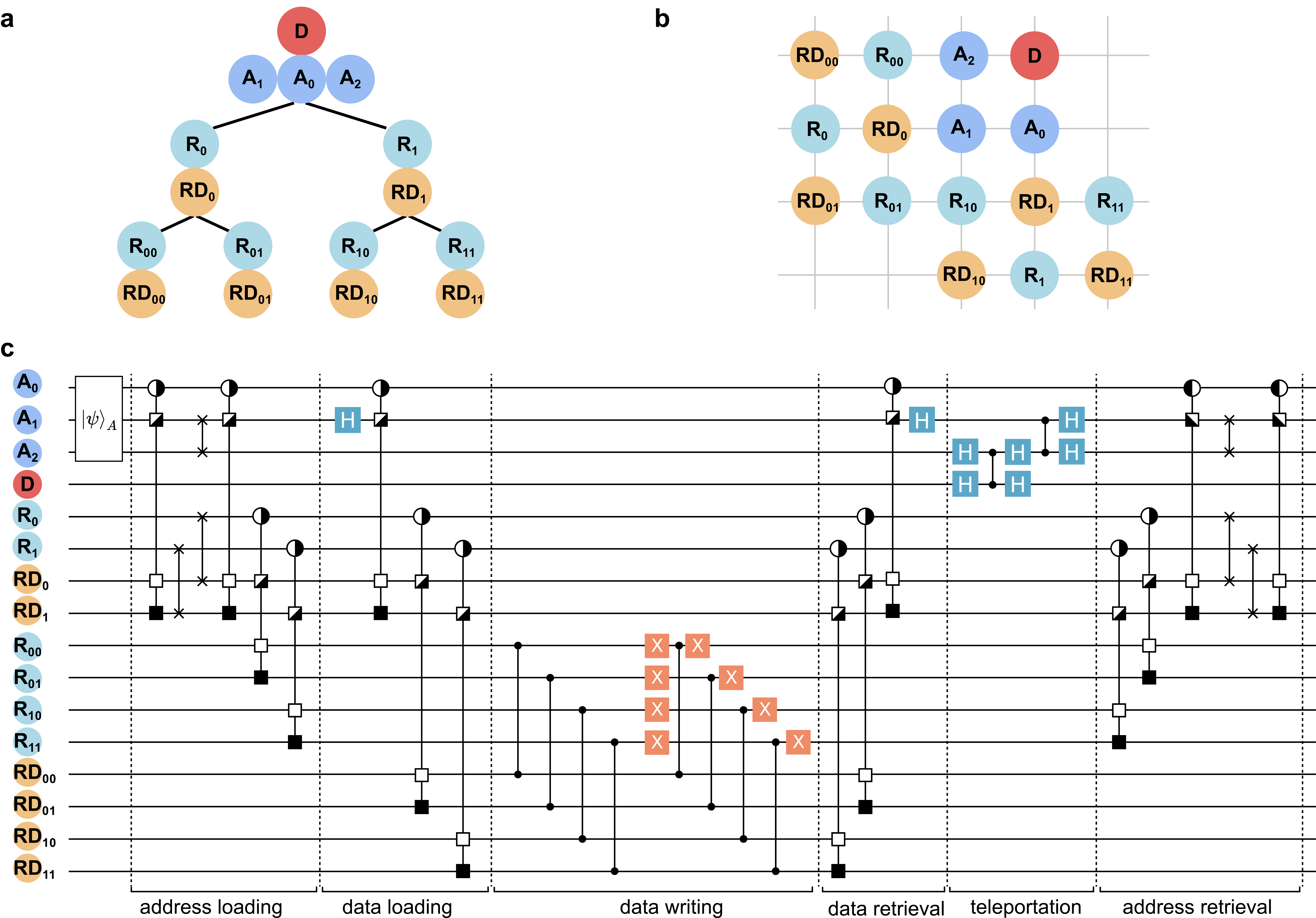}
    \caption{\textbf{QRAM structure and qubit layout.} In all figures, $A_i$ represents the qubits in the address register, $D$ represents the bits in the data register, and $R_i$ and $RD_i$ form a complete router, where $R_i$ is used to temporarily store the transmitted address. \textbf{a}, Three-layer bucket-brigade QRAM. \textbf{b}, Qubit layout of three-layer bucket-brigade QRAM. \textbf{c}, Experimental circuit of three-layer bucket-brigade QRAM.
    }
    \label{struct}
\end{figure*}

In Fig.1d of the main text, we present the experimental qubit layout for the two-layer bucket-brigade QRAM that utilized quantum teleportation. In this section, we will introduce the experimental qubit layout for the QRAM implementation of a three-layer bucket-brigade architecture. 
In \autoref{struct}, we present our simplified QRAM structure and the qubit layout in experiment.

\clearpage

\bibliography{ref}